\documentclass[11pt]{elsarticle}

\makeatletter
\def\ps@pprintTitle{%
 \let\@oddhead\@empty
 \let\@evenhead\@empty
 \def\@oddfoot{}%
 \let\@evenfoot\@oddfoot}
\usepackage{bm}
\usepackage{bbm}
\usepackage{amsmath}
\usepackage{amsthm}
\usepackage{empheq}
\usepackage{float}
\usepackage{indentfirst}
\usepackage{cases}
\usepackage{algorithm}  
\usepackage{algorithmicx} 
\usepackage{algpseudocode} 
\usepackage{subcaption}
\usepackage{graphicx}
\usepackage{booktabs}
\usepackage{diagbox}
\usepackage[all]{xy}
\usepackage{indentfirst}
\usepackage{epstopdf}
\usepackage{epsfig}
\usepackage{overpic}
\usepackage{array}
\usepackage{color}
\usepackage{multirow}
\usepackage[mathscr]{euscript}
\usepackage{latexsym,bm}
\usepackage{tikz}
\usetikzlibrary{shapes,arrows}
\usepackage[left=2.5cm,right=2.5cm,top=2cm,bottom=2cm]{geometry}
\usepackage{enumerate}

\usepackage{ulem}

\usepackage{framed,multirow}

\usepackage{amssymb}
\usepackage{latexsym}
\usepackage{mathtools}

\usepackage{url}
\usepackage{xcolor}
\definecolor{newcolor}{rgb}{.8,.349,.1}

\renewcommand{\vec}[1]{\boldsymbol{#1}} 

\newtheorem{remark}{Remark}
\def\bx{\bm{x}}
\def\bv{\bm{v}}

\def\D{\textup{d}}
\def\mone{\mathbbm{1}}

\makeatletter

\newcommand{\Rmnum}[1]{\expandafter@slowromancap\romannumeral #1@}
\makeatother

\tikzset{global scale/.style={
		scale=#1,
		every node/.append style={scale=#1}
	}
}
\tikzstyle{startstop} = [rectangle,rounded corners, minimum width=3cm,minimum height=1cm,text centered, draw=black,fill=red!30]
\tikzstyle{io} = [trapezium, trapezium left angle = 70,trapezium right angle=110,minimum width=3cm,minimum height=1cm,text centered,draw=black,fill=blue!30]
\tikzstyle{process} = [rectangle,minimum width=3cm,minimum height=1cm,text centered,text width =3cm,draw=black,fill=orange!30]
\tikzstyle{decision} = [diamond,minimum width=1.7cm,minimum height=0.5cm,shape aspect=2, text centered,draw=black,fill=green!30]
\tikzstyle{arrow} = [thick,->,>=stealth]

\begin{document}

\begin{frontmatter}

\title{\LARGE Adaptive sampling accelerates the hybrid deviational particle simulations}

\author[PKU]{Zhengyang Lei}
\ead{leizy@stu.pku.edu.cn}

\author[PKU]{Sihong Shao}
\ead{sihong@math.pku.edu.cn}


\address[PKU]{CAPT, LMAM and School of Mathematical Sciences,  Peking University, Beijing 100871, China}

\begin{abstract}
To avoid ineffective collisions between the equilibrium states, the hybrid method with deviational particles (HDP) has been proposed to integrate the Vlasov-Poisson-Landau system, while leaving a new issue in sampling deviational particles from the high-dimensional source term. 
In this paper, we present an adaptive sampling (AS) strategy that first adaptively reconstructs a piecewise constant approximation of the source term
based on sequential clustering via discrepancy estimation, and then samples deviational particles directly from the resulting adaptive piecewise constant function without rejection.  The mixture discrepancy, which can be easily calculated thanks to its explicit analytical expression, is employed as a measure of uniformity instead of the star discrepancy the calculation of which is NP-hard.  The resulting method, dubbed the HDP-AS method, samples deviational particles through adaptive sampling instead of the acceptance-rejection method in the original HDP method. In the Landau damping, two stream instability, bump on tail and Rosenbluth's test problems, the HDP-AS method runs
approximately ten times faster than the HDP method while keeping the same accuracy.

\end{abstract}


\end{frontmatter}

\section{Introduction}
\label{sec:intro}

\par
\begin{figure}[htb]
	\centering
	\includegraphics[width=0.45\textwidth]{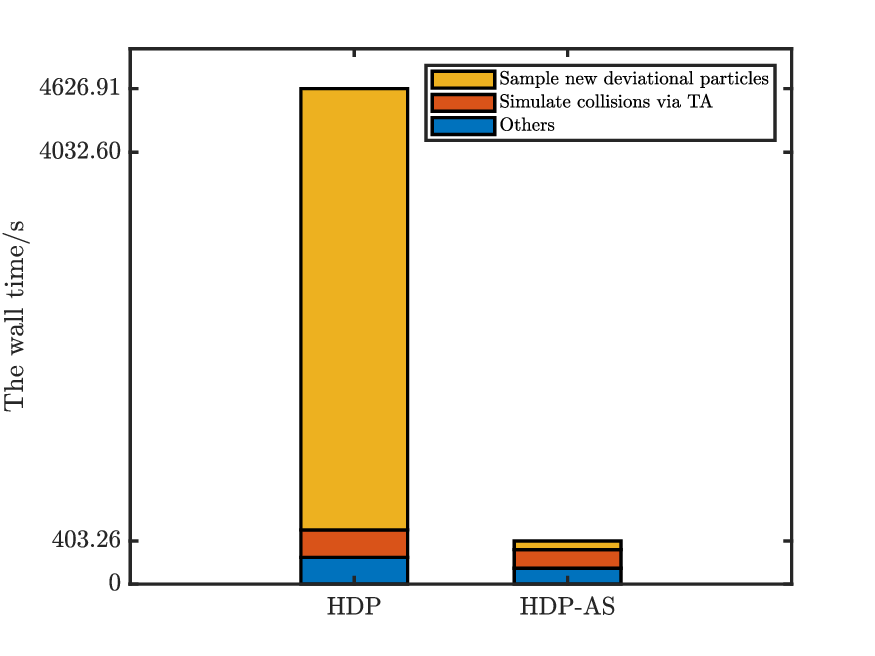}
	\caption{ The wall time consumed by sampling new deviational particles, simulating collisions via TA and others 	
		for HDP and HDP-AS in the linear Landau damping corresponding to Figure~\ref{0.01dist}. We find that sampling new deviational particles takes up only  22\% (89.91s/403.26s) of the total wall time  in HDP-AS,  whereas nearly 90\% (4120.41s/ 4629.91s) of the total wall time in HDP.}
	\label{ratio}
\end{figure}

 As a fundamental mathematical model for the kinetic theory, the Vlasov–Poisson–Landau (VPL) system 
\begin{equation}\label{VPL}
	\begin{aligned}
		\partial_t f + \vec{v} \cdot \nabla_{\vec{x}} f - \vec{E}\cdot \nabla_{\vec{v}} f &= Q(f,f), \\
		-\nabla_{\bx} \cdot \vec{E} &= \rho(\bx,t),  
	\end{aligned}
\end{equation}
describes the evolution of the charged particles in phase space $(\vec{x},\vec{v}) \in \mathbb{R}^3\times \mathbb{R}^3$ \cite{cercignani1988boltzmann, villani2002review}, where $f=f(\vec{x},\vec{v},t)$ gives the time-dependent distribution function, $\rho(\bx,t) = \int_{\mathbb{R}^3} f(\vec{x},\vec{v},t) \D \bv$ the spatial density and $Q(f,f)$ the Landau collision operator. The form of the VPL system \eqref{VPL} adopted in this paper is consistent with that presented in \cite{Yan2016}. Numerical resolution of the VPL system encounters some fundamental obstacles due to the notorious curse of dimensionality and thus particle-based methods play a prevalent role benefiting from their (possible) dimension independence.
The DSMC-PIC method, an often-used particle method, which exploits the operator splitting technique and combines the direct simulation Monte Carlo (DSMC) (such as the TA method by Takizuka and Abe \cite{Takizuka1977}  and the Nanbu method \cite{nanbu1997theory}) for the collision step
\begin{equation}\label{collision step}
	\partial_t f =Q(f, f), 
\end{equation}
and the particle in cell (PIC) \cite{Birdsall1991pic, grigoryev2012pic} for the advection step
\begin{equation}\label{vlasov poisson}
	\begin{aligned}
		\partial_t f + \vec{v} \cdot \nabla_{\vec{x}} f - \vec{E}\cdot \nabla_{\vec{v}} f &= 0, \\
		-\nabla_{\bx} \cdot \vec{E} &= \rho(\bx,t),
	\end{aligned}
\end{equation}
has been widely recognized in the literature. Accordingly,  a key challenge arises in the collision step \eqref{collision step} due to the high-dimensionality and nonlinearity of  $Q(f,f)$. The DSMC may become inefficient when the distribution function $f(\vec{x},\vec{v},t)$ approaches the equilibrium state, denoted by $m(\vec{x},\vec{v},t)$ (see Eq.~\eqref{maxwell}),  as the collisions between equilibrium parts satisfy $Q(m,m) \equiv 0$ which results in no net effect. 
To address this, a class of hybrid methods has been proposed to improve the efficiency via explicitly avoiding sampling from the ineffective collisions $Q(m,m)$ by the decomposition 
\[
f(\bx,\bv,t) = m(\bx,\bv,t) + f_d(\bx,\bv,t),
\] 
e.g., the micro-macro approach for the BGK operator \cite{crestetto2012kinetic}, the thermalization/dethermalization method for the spatial homogeneous Coulomb collisions \cite{caflisch2008hybrid} and the low-variance deviational simulation Monte Carlo methods for rarefied gas \cite{baker2005variance, Hadjiconstantinou2007low}. For the VPL system \eqref{VPL}, a hybrid method with deviational particles (HDP) was proposed along this line and allows the deviation $f_d$ to be negative \cite{Yan2015, Yan2016}. Namely, the HDP method decomposes the distribution as follows 
\begin{equation}\label{f decom}
	f(\bx,\bv,t) = m(\bx,\bv,t) + f_p(\bx,\bv,t) - f_n(\bx,\bv,t),
\end{equation}
where $f_p(\bx,\bv,t)\geq 0$ and $f_n(\bx,\bv,t)\geq 0$ denote the positive part and negative part of the difference $f(\bx,\bv,t)-m(\bx,\bv,t)$,  respectively. 
After that, how to sample new deviational particles from the high-dimensional source term $Q(f_p-f_n,m)$ at each collision step (see Eq.~\eqref{eqs decom}) becomes a new 
issue. The source term is a 5-D integral with singularity. HDP ingeniously approximates it as a 1-D integral and samples new deviational particles from it using the acceptance-rejection method \cite{Yan2015}. Nevertheless, this part remains very time-consuming in actual numerical simulations as a result of the low acceptance rate (see Figure~\ref{linear landau acc rate}) and the time spent on it accounts for, e.g.,  roughly 90\% of the total wall time (see Figure~\ref{ratio}) in the linear Landau damping. Actually, high-dimensional sampling poses challenges in scientific computing. The common methods, including the acceptance-rejection method and Markov Chain Monte Carlo method \cite{weinan2019applied}, can be used to sample from general distribution. However, it is hard to find a high acceptance rate upper bound that is convenient to sample for the acceptance-rejection method and the Markov Chain Monte Carlo method may get stuck in configuration space when the proposal variance is too high \cite{Roberts2001metropolis}. This work aims to solve this high-dimensional sampling problem from  $Q(f_p-f_n,m)$ at the collision step. 

\begin{figure}[htb]
	\centering
	\begin{subfigure}[b]{0.45\textwidth}
		\centering
		\includegraphics[width=\linewidth]{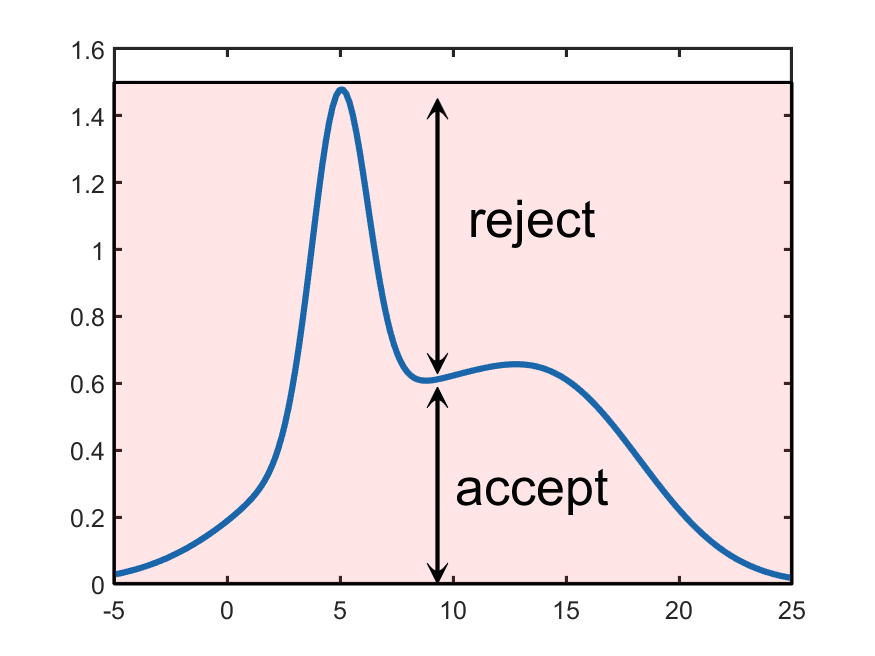}
		\caption{HDP: acceptance-rejection sampling.}
	\end{subfigure}
	\hfill
	\begin{subfigure}[b]{0.45\textwidth}
		\centering
		\includegraphics[width=\linewidth]{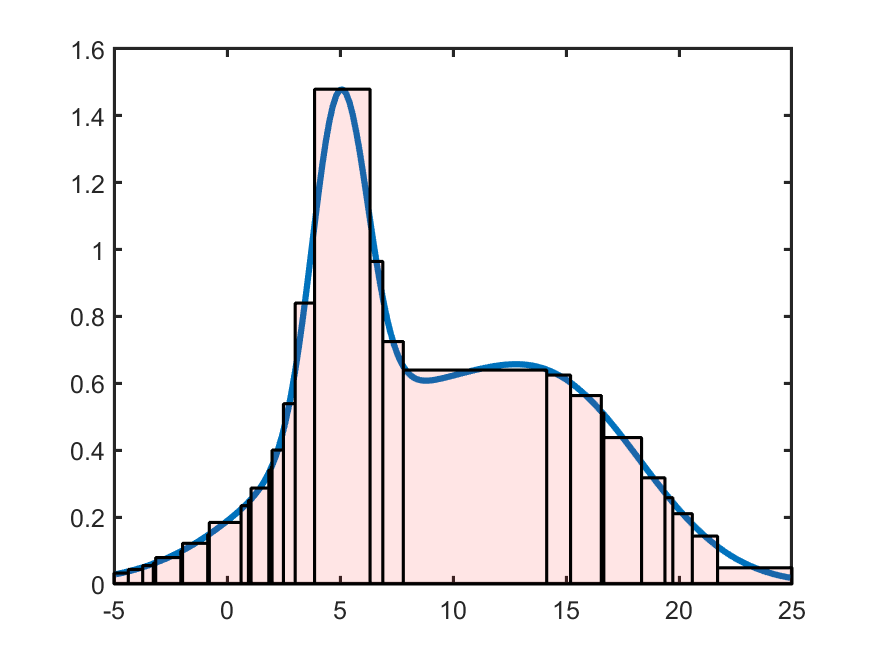}
		\caption{HDP-AS: adaptive sampling.}
	\end{subfigure}
	\caption{One-dimensional schematic diagram: The HDP-AS method samples new particles from an adaptive piecewise constant approximation without rejection while the HDP method employs the acceptance-rejection sampling \cite{Yan2015, Yan2016}.}
	\label{fig piecewise accrej}
\end{figure}

\par We propose an HDP method based on adaptive sampling (AS) for solving the VPL system \eqref{VPL}. The resulting method, dubbed the HDP-AS method, reconstructs an adaptive piecewise constant approximation of $Q(f_p-f_n,m)$ from the point distribution, which can be readily sampled from (see Figure~\ref{fig piecewise accrej} for a quick demonstration), and the wall time consumed by sampling new deviational particles is significantly reduced (see Figure~\ref{ratio}).
This two-phase architecture --- first constructing a piecewise constant approximation of the target distribution and then sampling from the resulting piecewise constant approximation --- 
can trace its lineage to the 1960s ziggurat algorithm \cite{Marsaglia1961, Marsaglia2000}
and dramatically enhance  the sampling efficiency. Moreover, HDP-AS partially borrows the idea from the discrepancy-based density estimation \cite{Li2016}, which employs the star discrepancy to measure irregularity of points distribution and then constructs an adaptive approximation of high-dimensional unknown function.
Considering the fact that the complexity of calculating the star discrepancy is NP-hard \cite{gnewuch2012new} and a high computational cost may be inevitable for a large sample sizes, HDP-AS replaces the star discrepancy with the mixture discrepancy which can be calculated explicitly \cite{zhou2013mixture}. 
The only difference between HDP and HDP-AS lies in that HDP samples new deviational particles by the acceptance-rejection method, whereas HDP-AS employs adaptive sampling. The other steps remain the same for both methods.
Numerical results for the Landau damping, two stream instability, bump on tail and Rosenbluth's test problem demonstrates that HDP-AS achieves about ten times acceleration compared with HDP while maintaining the same accuracy.

\par The rest of this paper is organized as follows. In Section~\ref{sec:background}, we first present some preliminaries of the VPL system, and then introduce related numerical attempts, including the TA method and the HDP method, which lay the groundwork for the proposed HDP-AS method. Section~\ref{sec:as} details the HDP-AS method. Numerical experiments are conducted in Section~\ref{sec:num} for comparing HDP-AS with HDP. The paper is concluded in Section \ref{sec:con}. 
This paper addresses sampling new deviational particles from the source term in the collision step of the HDP method. Other components like the advection step’s macro-micro decomposition and collision step’s particle annihilation remain identical between HDP-AS and HDP and are thus omitted here for brevity. 
The interested readers can refer to \cite{Yan2015, Yan2016} for more details.

\section{Preliminaries and background}
\label{sec:background}

\par The Landau collision operator in Eq.~\eqref{VPL}  modules the binary collisions governed by the long range Coulomb interaction and has the form
\begin{equation*}\label{LFPoperator}
		Q(f, f)(\bx,\bv,t)=\nabla_{\bv}\cdot  \int_{\mathbb{R}^3} \Phi\left(\bv-\bv^*\right)[\nabla_{\bv} f(\bx,\bv,t) f\left(\bx,\bv^*,t\right)
		-\nabla_{\bv^*} f\left(\bx,\bv^*,t\right) f(\bx,\bv,t)] \D \bv^*,
\end{equation*}
where $\Phi(\bv)=\frac{\vec{I}}{|\bv|}-\frac{\bv \otimes \bv}{|\bv|^3}$ with $\vec{I}$ being the $3\times 3$ identity matrix. 
For the equilibrium state given by the Maxwellian form
\begin{equation}\label{maxwell}
	m(\bx, \bv, t)=\frac{\rho(\bx, t)}{(2 \pi T(\bx,t))^{3 / 2}} \exp\left(-\frac{|\bv-\vec{u}(\bx,t)|^2}{2 T(\bx,t)}\right),
\end{equation}
it can be easily verified that $Q(m,m) \equiv 0$, where the macroscopic velocity $\vec{u}$ and temperature $T$ are defined by
\begin{equation}\label{marco}
	\vec{u}=\frac{1}{\rho} \int_{\mathbb{R}^3} f \bv \D\bv, \quad T=\frac{1}{3 \rho} \int_{\mathbb{R}^3} f|\bv-\vec{u}|^2 \D \bv .
\end{equation}

\par A hybrid representation of the distribution function, which involves both grids and particles,  is widely used to integrate the VPL system \eqref{VPL}. In the spatial direction, uniform grids $\mathcal{C}_k,\,k = 1,\dots,N_x$ are constructed with $\bx_k$ being the center point
of $\mathcal{C}_k$. The macroscopic quantities $\rho$, $\vec{u}$ and $T$ depend only on the position, and thus are defined on the spatial grids. 
In the velocity direction, the particle representation is employed. The positive part $f_p$ and negative part $f_n$ are represented by particles called P-particles and N-particles respectively. 
Denote the sets of these particles at $\bx_k$ and time $t$ as $\mathbb{P}_k(t)$ and  $\mathbb{N}_k(t)$, the size of which is $N_{p,k}$ and $N_{n,k}$, respectively.

To solve the VPL system \eqref{VPL}, we also start with an operator splitting between the collision part \eqref{collision step} and the advection part \eqref{vlasov poisson}. 
In the HDP-AS method, the simulation of advection part employs the identical micro-macro decomposition framework \cite{bennoune2008uniformly} as that utilized in HDP, so it will not be repeated here. Below we illustrate how the TA \cite{Takizuka1977} and HDP \cite{Yan2016} methods simulate collisions in one time step $\Delta t$, which lay the groundwork for the proposed HDP-AS method.

\subsection{The TA method}\label{TA sec}


\par The Landau operator $Q(g,h)$ describes the changes in $h$ resulting from collisions with $g$ and we refer to it as G-H collisions. The particles with the distribution $g$ and that with the distribution $h$ are respectively named as G-particles and H-particles. Denote the sets of G-particles and H-particles at grid point $\bx_k$ and time $t$ as $\mathbb{G}_k(t)$ and $\mathbb{H}_k(t)$, whose size is $N_{g,k}$ and $N_{h,k}$, respectively.
It is required that $N_{g,k}\geq N_{h,k}$. The TA method randomly matches the H-particles with G-particles to simulate binary collisions. 
Suppose the current velocities of a certain pair of particles are $\bv^{g}$ and $\bv^{h}$, and the velocities after the collision are $\widetilde{\bv}^{g}$ and $\widetilde{\bv}^{h}$. When the particle masses are equal, the conservation of momentum and energy yields
\begin{align}
	\bv^{g} +   \bv^{h} &=   \widetilde{\bv}^{g} +   \widetilde{\bv}^{h}, \label{Conservation of momentum}\\
	|\bv^{g}|^2 +   |\bv^{h}|^2 &=   |\widetilde{\bv}^{g}|^2 +   |\widetilde{\bv}^{h}|^2,\label{Conservation of power}
\end{align}
and thus 
\begin{equation}\label{after coll}
	\begin{aligned}
		\widetilde{\bv}^{g} &= \bv^{g} + \frac{1}{2} \Delta \vec{u}, \\
		\widetilde{\bv}^{h} &= \bv^{h} - \frac{1}{2} \Delta \vec{u}, \\
	\end{aligned}
\end{equation}
where $\Delta \vec{u} = (\widetilde{\bv}^{h}-\widetilde{\bv}^{g})-(\bv^{h}-\bv^{g})$ is the change of relative velocity. During the collision process, the magnitude of the relative velocity remains unchanged, but its direction is altered by the scattering angle $\Theta$ and azimuthal angle $\Phi$, which characterize the nature of the Coulomb collisions. In the TA method,  $\Phi$ is chosen randomly with a uniform distribution in $\left[0,2\pi\right]$ and $\Theta = 2 \arctan(\delta)$,
where $\delta$ is sampled from the normal distribution $\mathcal{N}(0,C\Delta t)$ and $C$ is determined by some physical constants as well as the magnitude of the relative velocity (see \cite{Takizuka1977} for more details). Let $\bv^{h}-\bv^{g} := (u_x,u_y,u_z)$ and $u := \sqrt{u_x^2+u_y^2+u_z^2}$. The change of relative velocity $\Delta \vec{u} := (\Delta u_x, \Delta u_y, \Delta u_z)$ can be computed by
\begin{equation}\label{relative vec}
	\begin{aligned}
		\Delta u_{x}&=\frac{u_x}{\sqrt{u_x^2+u_y^2}} u_{z} \sin \Theta \cos \Phi-\frac{u_y}{\sqrt{u_x^2+u_y^2}} u \sin \Theta \sin \Phi-u_{x}(1-\cos \Theta), \\
		\Delta u_{y}&=\frac{u_x}{\sqrt{u_x^2+u_y^2}} u_{z} \sin \Theta \cos \Phi+\frac{u_x}{\sqrt{u_x^2+u_y^2}} u \sin \Theta \sin \Phi-u_{y}(1-\cos \Theta), \\
		\Delta u_{z}&=-\sqrt{u_x^2+u_y^2} \sin \Theta \cos \Phi-u_{z}(1-\cos \Theta).
	\end{aligned}
\end{equation}
Once $\Delta \vec{u}$ is determined, the velocities after the collision $\widetilde{\bv}^{g}$ and $\widetilde{\bv}^{h}$ can be obtained according to Eq.~\eqref{after coll}. Algorithm \ref{TA alg} presents the pseudo-code of the TA method.

\begin{algorithm}[htbp]
	\caption{Simulate the G-H collisions with the TA method} 
	\hspace*{0.02in} {\bf Input:}
	\label{TA alg}
	Grid number $N_x$, the set of G-particles $\mathbb{G}_k(t)$ and the set of H-particles $\mathbb{H}_k(t)$, $k = 1,\dots,N_x$. \\
	\hspace*{0.02in} {\bf Output:}  The set of H-particles $\mathbb{H}_k(t+\Delta t)$ after the Coulomb collisions, $k = 1,\dots,N_x$.
	\begin{algorithmic}[1]
		\For{$k = 1:N_x$}
		\State $\mathbb{H}_k(t+\Delta t) \leftarrow$ an empty set;
		\State $N_{h,k} \leftarrow$ the size of $\mathbb{H}_k(t)$;
		\State Generate a random permutation vector $(\sigma(1),\dots,\sigma(N_{h,k}))$ of $(1,\dots,N_{h,k})$;
		\For{$i=1:N_{h,k}$}
		\State $\bv^{g} \leftarrow$ the velocity of the $\sigma(i)$-th G-particles;
		\State $\bv^{h}\leftarrow$ the velocity of the $i$-th H-particles;
		\State Generate randomly the scattering angle $\Theta$ and azimuthal angle $\Phi$;
		\State Compute the change of relative velocity $\Delta \vec{u}$ according to Eq.~\eqref{relative vec};
		\State Compute the velocity after the collision $\widetilde{\bv}^{h}$ according to Eq.~\eqref{after coll};
		\State Append $\widetilde{\bv}^{h}$ to the set $\mathbb{H}_k(t+\Delta t)$;
		\EndFor
		\EndFor
	\end{algorithmic}
\end{algorithm}

\subsection{The HDP method}

\par The TA method may become inefficient when the distribution $f(\bx,\bv,t)$ approaches the equilibrium state $m(\bx,\bv,t)$, as the collisions between equilibrium parts do not produce actual effect. According to the decomposition \eqref{f decom}, the HDP method solves the following system of equations \cite{Yan2015, Yan2016}
\begin{equation}\label{eqs decom}
	\left\{\begin{array}{l}
		\partial_t \tilde{f}=Q(\tilde{f}, \tilde{f}), \\
		\partial_t m=0, \\
		\partial_t f_p=Q(\tilde{f}, f_p)+\left(Q\left(f_{p}, m\right)-Q\left(f_{n}, m\right)\right)_{+}, \\
		\partial_t f_n=Q(\tilde{f}, f_n)+\left(Q\left(f_{p}, m\right)-Q\left(f_{n}, m\right)\right)_{-},
	\end{array}\right.
\end{equation}
explicitly removing the collisions $Q(m,m)$ of the equilibrium parts, where $h_{\pm} = \frac{1}{2}\left(|h|\pm h\right)$ gives the positive/negative part. Here the notation $\tilde{f}$ is used to distinguish it from the desired solution $f = m+f_p-f_n$ in Eq.~\eqref{f decom}. It is introduced in HDP as a coarse approximation of $f$ and used in simulating collisions  $Q(\tilde{f}, f_p)$ and $Q(\tilde{f}, f_n)$. The distribution $\tilde{f}$ is represented by F-particles and the set of these particles at $\bx_k$ and time $t$ is denoted as $\mathbb{F}_k(t)$, the size of which is $N_{f,k}$. The three parts $\tilde{f}, f_p, f_n$ are presented respectively by the F-particles, P-particles and N-particles, i.e., $\forall\,\varphi(\vec v)\in \mathbb{H}_0^2\left(\mathbb{R}^3\right)$, $\forall\, k \in\{ 1,\dots,N_x\}$,
\begin{equation}\label{fpn weak}
	\begin{aligned}
		\langle\varphi, \tilde{f}(\bx_k,\bv,t)\rangle 
		&\approx N_{\text {eff }}^F \sum_{\bv_{\tilde{f}} \in \mathbb{F}_k(t)}  \varphi(\bv_{\tilde{f}}), \\
		\left\langle\varphi, f_p(\bx_k,\bv,t)\right\rangle 
		&\approx N_{\text {eff }} \sum_{\bv_{f_p} \in \mathbb{P}_k(t)}  \varphi(\bv_{f_p}), \\
		\left\langle\varphi, f_n(\bx_k,\bv,t)\right\rangle 
		&\approx N_{\text {eff }} \sum_{\bv_{f_n} \in \mathbb{N}_k(t)}  \varphi(\bv_{f_n}), 
	\end{aligned}
\end{equation}
where the standard inner product $\langle g, h\rangle=\int_{\mathbb{R}^3} g h \D \bv$ is adopted. The effective number $N_{\text {eff }}^F$ and $N_{\text {eff }}$ represent the physical densities per numerical particle and satisfy
\begin{equation}\label{eff num}
	\begin{aligned}
		N_{f,k}N_{\text {eff }}^F \approx \rho(\bx_k,t) = \int_{\mathbb{R}^3} \tilde{f}(\bx_k,\bv,t) \D \bv,\\
		 N_{p,k} N_{\text {eff }}\approx \rho_p(\bx_k,t) = \int_{\mathbb{R}^3} f_p(\bx_k,\bv,t) \D \bv, \\
		  N_{n,k} N_{\text {eff }}\approx \rho_n(\bx_k,t) = \int_{\mathbb{R}^3} f_n(\bx_k,\bv,t) \D \bv.
	\end{aligned}
\end{equation}
Eq.~\eqref{eff num} can be obtained by taking $\varphi \equiv 1$ in Eq.~\eqref{fpn weak}. 
A key ingredient is that F-particles are not necessary to be a very accurate approximation of $f$, and one only needs $N_{f,k}\geq N_{p,k}+N_{n,k}$ to ensure there are enough F-particles to simulate collisions $Q(\tilde{f}, f_p)$ and $Q(\tilde{f}, f_n)$. An error analysis shows that F-particles introduces a negligible error (see Section~3.3 in \cite{Yan2015}). Eq.~\eqref{eqs decom} guides the evolution of the F-particles, P-particles and N-particles in 
a prediction–correction manner.
The F-particles act as a prediction of $f$, and they are still simulated by the TA method. 
As corrections, P-particles and N-particles contain two parts in their evolution: One is collisions with F particles, simulated using the TA method; the other is sampling new particles from the source term
\begin{equation}\label{eq:mk}
	\Delta m_k(\bv) := \Delta t\left(Q\left(f_p(\bx_k,\bv,t), m(\bx_k,\bv,t)\right)-Q\left(f_n(\bx_k,\bv,t), m(\bx_k,\bv,t)\right)\right).
\end{equation} 
Algorithm \ref{HDP alg} exhibits the pseudo-code of the HDP method. We focus on the main steps of HDP. There are some other steps (such as enforcing conservation and particle resampling to alleviate the growth of the particle number \cite{Yan2015, Yan2016}) for which HDP-AS and HDP are consistent and are not listed in Algorithm \ref{HDP alg} for brevity. 

\begin{algorithm}[htbp]
	\caption{Simulate the Coulomb collisions with the HDP method} 
	\hspace*{0.02in} {\bf Input:}
	\label{HDP alg}
	Grid number $N_x$, the set of F-particles $\mathbb{F}_k(t)$, the set of P-particles $\mathbb{P}_k(t)$ and the set of N-particles $\mathbb{N}_k(t)$, $k = 1,\dots,N_x$. \\
	\hspace*{0.02in} {\bf Output:} The set of F-particles $\mathbb{F}_k(t+\Delta t)$, P-particles $\mathbb{P}_k(t+\Delta t)$ and N-particles $\mathbb{N}_k(t+\Delta t)$ after the Coulomb collisions, $k = 1,\dots,N_x$.
	\begin{algorithmic}[1]
		\State Simulate F-F collisions via the TA method in Algorithm~\ref{TA alg} and obtain $\mathbb{F}_k(t+\Delta t)$, $k = 1,\dots,N_x$;
		\State Simulate F-P collisions via the TA method in Algorithm~\ref{TA alg} and obtain $\mathbb{P}_k(t+\Delta t)$, $k = 1,\dots,N_x$;
		\State Simulate F-N collisions via the TA method in Algorithm~\ref{TA alg} and obtain $\mathbb{N}_k(t+\Delta t)$, $k = 1,\dots,N_x$;
		\For{$k=1:N_x$}
		\State Sample new P-particles from $\Delta m_k(\bv)_{+}$ and add them to $\mathbb{P}_k(t+\Delta t)$;
		\State Sample new N-particles from $\Delta m_k(\bv)_{-}$ and add them to $\mathbb{N}_k(t+\Delta t)$;
		\EndFor
	\end{algorithmic}
\end{algorithm}


\par The acceptance-rejection sampling (see Figure~\ref{fig piecewise accrej}) was adopted to sample new P-particles and N-particles respectively from $\Delta m_k(\bv)_{+}$ and $\Delta m_k(\bv)_{-}$ required by Algorithm~\ref{HDP alg} \cite{Yan2015, Yan2016}. 
Suppose that $|\Delta m_k(\bv)|$ has an upper bound $\widehat{ \Delta m_k}(\bv)$ satisfying $| \Delta m_k(\bv)|\leq \widehat{ \Delta m_k}(\bv)$. One can sample particles from $\widehat{ \Delta m_k}(\bv) / \int_{\mathbb{R}^3} \widehat{ \Delta m_k}(\bv) \D\bv$. For one particle with coordinate $\bv_0$, accept it with probability $ |\Delta m_k(\bv_0)|/\widehat{ \Delta m_k}(\bv_0)$ and label it a P or N particle in accordance with the sign of $ \Delta m_k(\bv_0)$. There are two prerequisites for such acceptance-rejection sampling to be efficient: (a) The upper bound function $\widehat{ \Delta m_k}(\bv)$ is easy to sample, and (b) the acceptance probability $|\Delta m_k(\bv)|/\widehat{ \Delta m_k}(\bv)$ is high. However, (a) and (b) are usually not compatible. If $\Delta m_k(\bv)$ is a general function and $\widehat{ \Delta m_k}(\bv)$ is a readily sampled probability density, then the gap between them is usually relatively large, which will lead to the low acceptance probability $|\Delta m_k(\bv)|/\widehat{ \Delta m_k}(\bv)$. This also constitutes the main reason why we try to develop an adaptive sampling strategy for the HDP simulations.

\section{Adaptive sampling}
\label{sec:as}


\par Different from the acceptance-rejection sampling in the HDP method \cite{Yan2015, Yan2016}, HDP-AS proposes an adaptive high-dimensional sampling approach to sample from $\Delta m_k(\bv)_{\pm}$ required by Algorithm \ref{HDP alg}.
There exist two steps in the HDP-AS method: (a) Provide an adaptive piecewise constant approximation of the function $\Delta m_k(\bv)$, 
denoted by $\overline{ \Delta m_k}(\bv)$,  and (b) sample new deviational particles directly from $\overline{ \Delta m_k}(\bv)$
instead of $\Delta m_k(\bv)$. In contrast to the acceptance-rejection sampling, the implementation of sampling from a piecewise constant function in (b) is much more straightforward and without rejection. Moreover, we employ adaptive piecewise constants instead of uniform ones to alleviate the curse of dimensionality. As the implementation of (b) is straightforward, we focus exclusively on how to achieve (a) in an efficient manner in the following.

\par
It involves two processes: 
(a1) Simulate the Coulomb collisions $Q(f_p,m)$ and $Q(f_n,m)$ in the function $\Delta m_k(\bv)$ via the TA method in Algorithm~\ref{TA alg} and obtain the point distribution $W_k(\bv)$ (see Eq.~\eqref{Wk}), approximating $Q\left(f_{p}, m\right)-Q\left(f_{n}, m\right)$ in the weak sense, and (a2) adaptively reconstruct $\overline{ \Delta m_k}(\bv)$ from the point distribution $W_k(\bv)$.
Denote the particles with the equilibrium distribution $m$ as M-particles, which are not tracked and only sampled from \eqref{maxwell} whenever needed. In $\Delta m_k(\bv)$, $Q(f_p,m)$ and $Q(f_n,m)$ represent that P-M collisions and N-M collisions respectively, both of which can be simulated by the TA method. Let $\mathbb{M}^{+}_k$ and $\mathbb{M}^{-}_k$ collect all the M-particles after P-M collisions and N-M collisions at grid point $\bx_k$ and time $t$, respectively. They can form a point distribution
\begin{equation}\label{Wk}
	W_k(\bv)  = N_{\text {eff }} \sum_{\vec a\in \mathbb{M}^{+}_k} \delta(\bv-\vec a) - N_{\text {eff }}\sum_{\vec a\in \mathbb{M}^{-}_k} \delta(\bv-\vec a),
\end{equation}
where $\delta(\bv)$ is the Dirac distribution in $\mathbb{R}^3$. Actually, the TA method ensures that $W_k$ in Eq.~\eqref{Wk}
may approximate $Q\left(f_{p}, m\right)-Q\left(f_{n}, m\right)$ in the weak sense, i.e., $\forall\,\varphi(\vec v)\in \mathbb{H}_0^2\left(\mathbb{R}^3\right)$, $\forall\, k \in\{ 1,\dots,N_x\}$,
\begin{equation}\label{m+-}
	\left\langle\varphi, Q\left(f_p(\bx_k,\bv,t),m(\bx_k,\bv,t)\right)-Q\left(f_n(\bx_k,\bv,t), m(\bx_k,\bv,t)\right)\right\rangle 
	\approx \left\langle\varphi, W_k  \right\rangle
\end{equation}
with 
\begin{equation}
	\left\langle\varphi, W_k  \right\rangle = N_{\text {eff }} \sum_{\vec a\in \mathbb{M}^{+}_k} \varphi\left(\vec a\right)-N_{\text {eff }} \sum_{\vec a\in \mathbb{M}^{-}_k} \varphi\left(\vec a\right).
\end{equation}
Algorithm \ref{delta m alg} details the way of obtaining $\mathbb{M}^{\pm}_k$ in Eq.~\eqref{Wk}.

\begin{algorithm}[htbp]
	\caption{Simulate the Coulomb collisions in $\Delta m_k(\bv)$ for $k = 1,\dots,N_x$} 
	\hspace*{0.02in} {\bf Input:}
	\label{delta m alg}
	Grid number $N_x$, the set of P-particles $\mathbb{P}_k(t)$,  the set of N-particles $\mathbb{N}_k(t)$, the macroscopic velocity $\vec{u}(\bx_k,t)$ and temperature $T(\bx_k,t)$. \\
	\hspace*{0.02in} {\bf Output:}  Particle sets $\mathbb{M}^{\pm}_k$ in Eq.~\eqref{Wk}.
	\begin{algorithmic}[1]
		\For{$k=1:N_x$}
		\State $\mathbb{M}_k \leftarrow$  an empty set;
		\State $N_{p,k} \leftarrow$ the size of $\mathbb{P}_k(t)$;
		\State Sample $N_{p,k}$ particles from the Maxwellian distribution \eqref{maxwell} and add them to $\mathbb{M}_k$;
		\EndFor
		\State Simulate P-M collisions via the TA method in Algorithm~\ref{TA alg} and  obtain $\mathbb{M}^{+}_k$;
		\For{$k=1:N_x$}
		\State $\mathbb{M}_k \leftarrow$  an empty set;
		\State $N_{n,k} \leftarrow$ the size of $\mathbb{N}_k(t)$;
		\State Sample $N_{n,k}$ particles from the Maxwellian distribution \eqref{maxwell} and add them to $\mathbb{M}_k$;
		\EndFor
		\State Simulate N-M collisions via the TA method in Algorithm~\ref{TA alg} and obtain $\mathbb{M}^{-}_k$;
	\end{algorithmic}
\end{algorithm}

\par Next, we need to accomplish the process (a2). Starting from the weak approximation \eqref{m+-}, the HDP-AS method seeks the adaptive piecewise constant approximation $\overline{ \Delta m_k}(\bv)$ of $\Delta m_k(\bv)$. Let $\Omega = \left[a,b\right]^3 \subset \mathbbm{R}^3$ be the computational domain containing all particles in the sets $\mathbb{M}^{+}_k$ and $\mathbb{M}^{-}_k$, and it can be decomposed into 
\begin{equation}\label{domain dec}
	\Omega = \left[a,b\right]^3 = \bigcup_{l=1}^{L} \Omega_l,
\end{equation}
such that $\Delta m_k(\bv)$ can be approximated as a constant $c_l \Delta t$ in each sub-domain $\Omega_l$, namely,
\begin{equation}\label{ada pie con}
	\Delta m_k(\bv) \approx \overline{ \Delta m_k}(\bv) := \sum_{l=1}^{L} c_l \Delta t \mone_{\Omega_l}(\bv),
\end{equation}
where $L$, $c_l$ and $\Omega_l$ depends on the position $\bx_k$, but for brevity of notation, we omit the subscript $k$ representing such dependence. 
Taking the test function $\varphi$ as the indicator function $\mone_{\Omega_l}$ in Eq.~\eqref{m+-}, we have 
\begin{equation}\label{ck}
	\begin{aligned}
		c_l &\approx \frac{1}{\mu\left(\Omega_l\right)} \int_{\Omega_l} (Q\left(f_p(\bx_k,\bv,t),m(\bx_k,\bv,t)\right)-Q\left(f_n(\bx_k,\bv,t), m(\bx_k,\bv,t)\right)) \D \bv \\
		&\approx N_{\text {eff }}\sum_{\vec a\in \mathbb{M}^{+}_k} \frac{ \mone_{\Omega_l}\left(\vec a\right)}{\mu\left(\Omega_l\right)} - N_{\text {eff }} \sum_{\vec a\in \mathbb{M}^{-}_k} \frac{ \mone_{\Omega_l}\left(\vec a\right)}{\mu\left(\Omega_l\right)},
	\end{aligned}
\end{equation}
where $\mu\left(\Omega_l\right)$ gives the Lebesgue measure of $\Omega_l$. A simple idea for the domain decomposition in Eq.~\eqref{domain dec} is to utilize uniform grids, that is, $\Omega_l, \, l = 1\dots,L$ are hypercubes with the same side lengths. In spite of its simplicity and easy implementation, uniform grids tend to cause large errors. Small approximation error is achieved only if the side lengths of $\Omega_l$ are small, whereas uniform grids of small lengths may lead to few particles within the grids and thus an increase in stochastic error. HDP-AS replaces the uniform grids with adaptive ones, and the idea originates from \cite{Li2016}. It seeks an adaptive partition of $\Omega$ via a recursive binary splitting until the particles inside each sub-domain $\Omega_l$ are close to uniform distribution, which means that $\Delta m_k(\bv)$ can be approximately considered as a constant within $\Omega_l$. Algorithm~\ref{adaptive alg} details the process of this adaptive partitioning. Let $\mathcal{P}_L=\left\{\Omega_1, \ldots, \Omega_L\right\}$ be the collection of sub-domains whose union is $\Omega$. In the beginning, $\mathcal{P}_1 = \left\{\Omega_1\right\}$ and $\Omega_1 = \Omega$. At level $L+1$, $\mathcal{P}_{L+1}$ is generated by splitting one of the regions in $\mathcal{P}_{L}$ along one coordinate and then merging both resulting sub-domains with the remaining regions in $\mathcal{P}_{L}$.

\begin{algorithm}[h]
	\caption{ Reconstruct the adaptive piecewise constant approximation $\overline{ \Delta m_k}(\bv)$} 
	\hspace*{0.02in} {\bf Input:}
	\label{adaptive alg}
	The domain $\Omega$, particle sets $\mathbb{M}^{+}_k$ and $\mathbb{M}^{-}_k$, parameters $\vartheta, q$.\\
	\hspace*{0.02in} {\bf Output:} The piecewise constant function $\overline{ \Delta m_k}(\bv)$.
	\begin{algorithmic}[1]
		\State $L=1, \Omega_1=\Omega, \mathcal{P}_1=\left\{\Omega_1\right\}$;
		\State $N_{p,k} \leftarrow$ the size of $\mathbb{M}^{+}_k$;
		\State $N_{n,k} \leftarrow$ the size of $\mathbb{M}^{-}_k$;
		\State $N = N_{p,k} + N_{n,k}$; \label{l3}
		\While{true}
		\State flag = 0;
		\ForAll{$\Omega_l \in \mathcal{P}_L$}
		\State $\mathbb{M}^{+}_{\text{cur}} \leftarrow$ the subset of particles in $\mathbb{M}^{+}_k$ that fall within $\Omega_l$; \label{l1}
		\State $\mathbb{M}^{-}_{\text{cur}} \leftarrow$  the subset of particles in $\mathbb{M}^{-}_k$ that fall within $\Omega_l$; \label{l2}
		\State $N_{+} \leftarrow$ the size of $\mathbb{M}^{+}_{\text{cur}}$; \label{l4}
		\State $N_{-} \leftarrow$ the size of $\mathbb{M}^{-}_{\text{cur}}$; \label{l5}
		\State Calculate the discrepancies $D(\mathbb{M}^{+}_{\text{cur}})$ and $D(\mathbb{M}^{-}_{\text{cur}})$;
		\If{$D(\mathbb{M}^{+}_{\text{cur}}) > \frac{\vartheta \sqrt{N}}{\max(N_+,N_{-})}$ or $D(\mathbb{M}^{-}_{\text{cur}}) > \frac{\vartheta \sqrt{N}}{\max(N_+,N_{-})}$}
		\State Choose a splitting node $\widetilde{s}_{i_0, j_0}^{(l)}$ according to Eq.~\eqref{split c};
		\State Divide $\Omega_l$ into $\Omega_l^{(1)} \cup \Omega_l^{(2)}$ as Eq.~\eqref{split 12}; 
		\State $\Omega_l \leftarrow \Omega_l^{(1)}, Q_{L+1} \leftarrow \Omega_l^{(2)}$;
		\State flag = 1;
		\State Break;
		\EndIf
		\EndFor
		\If{flag = 1}
		\State $\mathcal{P}_{L+1}\leftarrow\left\{\Omega_1, \ldots, \Omega_{L+1}\right\}$;
		\State $L \leftarrow L+1$;
		\Else
		\State Break;
		\EndIf
		\EndWhile
		\State Calculate piecewise constants $c_l,\,l=1\dots,L$ according to Eq.~\eqref{ck};
		\State $\overline{ \Delta m_k}(\bv)\leftarrow\sum\limits_{l=1}^{L} c_l \Delta t \mone_{\Omega_l}(\bv)$;
	\end{algorithmic}
\end{algorithm}

Two key issues remains to be specified in the recursive binary splitting: Where to split and whether to split.
\begin{itemize}
	\item $\textbf{Where to split}$: Suppose $\Omega_l=\left[a_1^{(l)}, b_1^{(l)}\right] \times \left[a_2^{(l)}, b_2^{(l)}\right] \times\left[a_3^{(l)}, b_3^{(l)}\right]$ and $s_{i_0, j_0}^{(l)}=a_{j_0}^{(l)}+\frac{i_0}{q} \left(b_{i_0}^{(l)}-a_{i_0}^{(l)}\right)$ are some partition points to be selected where $i_0=1,2, \ldots,q-1$, $j_0=1,2,3$. To further divide the non-uniform sub-domains, the (sub)-optimal split points, denoted by $\widetilde{s}_{i_0, j_0}^{(l)}$, are suggested to maximize the difference gap
	\begin{equation}\label{split c}
		\widetilde{s}_{i_0, j_0}^{(l)}=\underset{s_{i_0, j_0}^{(l)}}{\arg \max }\left|\frac{N_{+}^{<}+N_{-}^{<}}{N_{+}+N_{-}} - \frac{i_0}{q}\right|,
	\end{equation}
	where $\Omega_l$ is divided into $\Omega_l^{(1)}$ and $\Omega_l^{(2)}$,
	\begin{equation}\label{split 12}
		\begin{aligned}
			& \Omega_l^{(1)}=\prod_{j=1}^{j_0-1}\left[a_j^{(l)}, b_j^{(l)}\right] \times\left[a_{j_0}^{(l)}, s_{i_0, j_0}^{(l)}\right] \times \prod_{j=j_0+1}^3\left[a_j^{(l)}, b_j^{(l)}\right], \\
			& \Omega_l^{(2)}=\Omega_l \backslash \Omega_l^{(1)},
		\end{aligned}
	\end{equation}
	and $N_{+}^{<}$ and $N_{-}^{<}$ count the number of particles in $\mathbb{M}^{+}_{\text{cur}}$ and $\mathbb{M}^{-}_{\text{cur}}$ (see Lines \ref{l1} and \ref{l2} in Algorithm \ref{adaptive alg}) that fall within $\Omega_l^{(1)}$, respectively.  The intuition behind Eq.~\eqref{split c} is to dig out the most non-uniform sub-domain after fixing the selectable partition points.
	\item $\textbf{Whether to split}$: We use the discrepancies $D(\mathbb{M}^{+}_{\text{cur}})$ and $D(\mathbb{M}^{-}_{\text{cur}})$ as indicators to measure the uniformity of the points distribution. Small discrepancies indicate a relatively uniform particle distribution and no further partitioning is necessary.  Discrepancy is a broad category and contains various specific definitions \cite{pausinger2015koksma, hickernell1998generalized}. The star discrepancy is used in \cite{Li2016}, the definition of which for point set $\left(\bv_1,\dots,\bv_n\right)\subset \left[0,1\right]^{3n}$ is stated as follows
	\begin{equation}
		D^*\left(\bv_1,\dots,\bv_n\right)=\sup _{\boldsymbol{u} \in[0,1]^{3}}\left|\frac{1}{n} \sum_{i=1}^n \mone_{[\mathbf{0}, \boldsymbol{u})}\left(\bv_i\right)-\operatorname{vol}([\mathbf{0}, \boldsymbol{u}))\right|.
	\end{equation}
	The complexity of calculating the star discrepancy is NP-hard. Although there are heuristic algorithms \cite{gnewuch2012new, Fang1997thres}, a high computational cost is likely to be unavoidable when the sample sizes are large. 
	To ensure efficient partitioning, we employ the mixture discrepancy (take $D$ in Algorithm \ref{adaptive alg} as $D^{\text{mix}}$ in Eq.~\eqref{mix discre}) to substitute the star discrepancy as the uniformity measure, which possesses an explicit expression \cite{zhou2013mixture}
	\begin{equation}\label{mix discre}
		\begin{aligned}
			D^{\text{mix}}\left(\bv_1,\dots,\bv_n\right)= & \left(\frac{19}{12}\right)^3-\frac{2}{n} \sum_{i=1}^n \prod_{j=1}^3\left(\frac{5}{3}-\frac{1}{4}\left|v_{i j}-\frac{1}{2}\right|-\frac{1}{4}\left|v_{i j}-\frac{1}{2}\right|^2\right) \\
			& +\frac{1}{n^2} \sum_{i=1}^n \sum_{k=1}^n \prod_{j=1}^3\left(\frac{15}{8}-\frac{1}{4}\left|v_{i j}-\frac{1}{2}\right|-\frac{1}{4}\left|v_{k j}-\frac{1}{2}\right|\right. \\
			& \left.-\frac{3}{4}\left|v_{i j}-v_{k j}\right|+\frac{1}{2}\left|v_{i j}-v_{k j}\right|^2\right),
		\end{aligned}
	\end{equation}
	where $\bv_i = (v_{i1},v_{i2},v_{i3})$. Let $\mathbb{M}^{+}_{\text{cur}} = \{\bv_1^+, \dots,\bv_{N_+}^+\}$, $\mathbb{M}^{-}_{\text{cur}} = \{\bv_1^-, \dots,\bv_{N_-}^-\}$. The discrepancy, for two sequences $\mathbb{M}^{+}_{\text{cur}}$ and $\mathbb{M}^{-}_{\text{cur}}$ in $\Omega_l$, can be defined by a linear scaling $\iota_l: \Omega_l \rightarrow \left[0,1\right]^{3}$,
	\begin{equation}
		\begin{aligned}
			D^{\text{mix}}(\mathbb{M}^{+}_{\text{cur}}) &= D^{\text{mix}}(\iota_l(\bv_1^+), \dots, \iota_l(\bv_{N_+}^+)), \\
			D^{\text{mix}}(\mathbb{M}^{-}_{\text{cur}}) &= D^{\text{mix}}(\iota_l(\bv_1^-), \dots, \iota_l(\bv_{N_-}^-)),
		\end{aligned}
	\end{equation}
	with 
	\begin{equation}
		\iota_l(v_1, v_2, v_3) = \left(\frac{v_1-a_1^{(l)}}{b_1^{(l)}-a_1^{(l)}}, \frac{v_2-a_2^{(l)}}{b_2^{(l)}-a_2^{(l)}}, \frac{v_3-a_3^{(l)}}{b_3^{(l)}-a_3^{(l)}}\right).
	\end{equation}
	The specified stopping criterion is
	\begin{equation}\label{stop criter}
		D^{\text{mix}}(\mathbb{M}^{+}_{\text{cur}}) \leq \frac{\vartheta \sqrt{N}}{\max(N_+,N_{-})},\quad D^{\text{mix}}(\mathbb{M}^{-}_{\text{cur}}) \leq \frac{\vartheta \sqrt{N}}{\max(N_+,N_{-})},
	\end{equation}
	where the parameter $\vartheta$ governs the depth of partition,
	and $N, N_\pm$ are given in Lines \ref{l3}, \ref{l4} and \ref{l5} in Algorithm \ref{adaptive alg}. A smaller $\vartheta$ leads to a finer partitioning. The above criterion~\eqref{stop criter} treats the P-particles and N-particles on the same footing and has been successfully applied into an efficient stochastic particle simulation of the Wigner-Coulomb quantum dynamics \cite{XiongShao2020Overcoming}.

\end{itemize}

\par  In contrast to the low acceptance rate of the acceptance-rejection method employed by HDP (see Figure \ref{linear landau acc rate}), sampling from the piecewise constant functions $\overline{ \Delta m_k}(\bv)$ in HDP-AS is straightforward and without rejection, thereby significantly enhancing the sampling efficiency. Using this improvement, the main time-consuming steps of HDP-AS become simulating collisions and advection, while sampling new deviational particles is no longer time-consuming now (see Figure~\ref{ratio}).
\begin{remark}
	The iconic ziggurat algorithm \cite{Marsaglia1961, Marsaglia2000} efficiently generates samples
	by partitioning the desired distribution functions into equal-area segments. Knuth praises it as ``a very pretty example of mathematical theory intimately interwoven with programming ingenuity---a fine illustration of the art of computer programming!'' in his famous book \cite{knuth1997}. This work inherits the idea of the ziggurat algorithm to solve the sampling problem in the HDP method. The main challenges in current applications lie in the desired distribution $\Delta m_k(\bv)$ is a three-dimensional function and we need efficiently construct a piecewise constant approximation from its point cloud form \eqref{Wk}. Our HDP-AS method addresses these specific challenges through an adaptive domain partitioning based on the mixture discrepancy.
\end{remark}

%
		%

%

\section{Numerical experiments}
\label{sec:num}


\par 
This section conducts five typical numerical experiments on the 1-D spatial and 3-D velocity coordinates,
including  the linear and nonlinear Landau damping problems, two stream instability problem, Bump on Tail problem and Rosenbluth’s problem.  A careful benchmark of HDP on the linear and nonlinear Landau damping against PIC-DSMC \cite{Yan2016}, which combines a Particle in Cell (PIC) method on the advection and a Direct Simulation Monte Carlo (DSMC) method on the collisions, as well as on Bump on Tail problem and Rosenbluth’s problem against the TA method \cite{Yan2015}, has been conducted and demonstrates the accuracy and efficiency of HDP. Therefore this section only focuses on a clear and one-to-one comparison between HDP-AS and HDP.  We set the spatial domain to be $x\in \left[0,4\pi\right]$ with periodic boundary condition and choose $N_x = 100$ grids with the spatial spacing $\Delta x = \frac{4\pi}{N_x}$. The time step is set as $\Delta t = \Delta x/10$. The partitioning parameters 
for the adaptive reconstruction in Algorithm~\ref{adaptive alg} are set as $\vartheta = 0.1$, $q = 16$ unless otherwise specified. We control the numerical particle numbers by the parameters $N_{\text {eff }}$ and $N_{\text {eff }}^F$.
In order to visualize the high-dimensional numerical solutions, we adopt the following 1-D and 2-D projections
\begin{equation}
	p_{1d}(v_1,t) = \int_{\mathbb{R}^3} f(x,\bv,t) ~\D x\D v_2 \D v_3,\quad  p_{2d}(x, v_1, t) = \int_{\mathbb{R}^2} f(x,\bv,t) ~\D v_2 \D v_3.
\end{equation}
Meanwhile, we use the relative $L^2$ error of the 2-D projection
\begin{equation}\label{err}
	\mathcal{E}_2[p_{2d}](t)=\frac{\left\|p_{2d}^{\mathrm{num}}(x, v_1, t)-p_{2d}^{\mathrm{ref}}(x, v_1,  t)\right\|_2}{\left\|p_{2d}^{\mathrm{ref}}(x, v_1, t)\right\|_2}
\end{equation}
to measure the accuracy. The reference solutions $p_{2d}^{\mathrm{ref}}(x, v_1, t)$ are produced by the DSMC-PIC method with $N_{\text {eff }}^F = 6.25\times 10^{-7}$. All simulations via C++ implementations run on the High-Performance Computing Platform of Peking University: 2*Intel Xeon E5-2697A-v4 (2.60GHz, 40MB Cache, 9.6GT/s QPI Speed, 16 Cores, 32 Threads) with 256GB Memory $\times$ 16.
\subsection{Linear Landau damping}
\begin{figure}[htbp]
	\centering
	\begin{subfigure}[b]{0.45\textwidth}
		\centering
		\includegraphics[width=\linewidth]{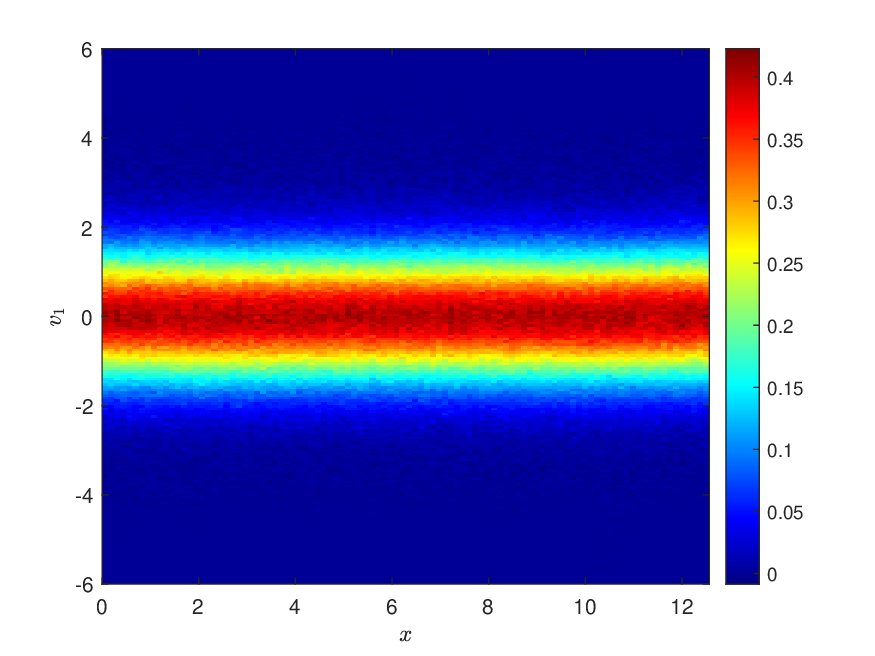}
		\caption{HDP.}
	\end{subfigure}
	\hfill
	\begin{subfigure}[b]{0.45\textwidth}
		\centering
		\includegraphics[width=\linewidth]{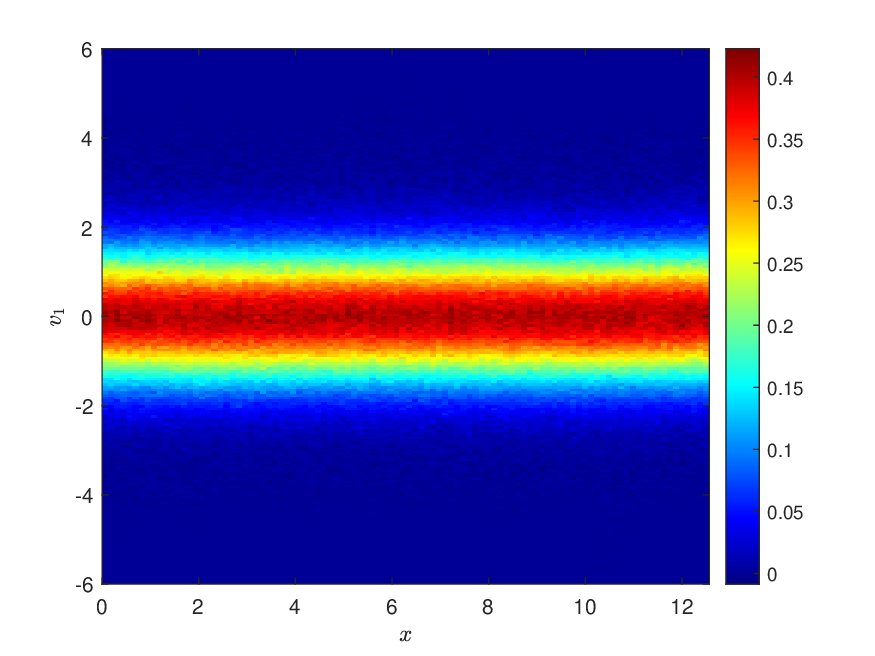}
		\caption{HDP-AS.}
	\end{subfigure}
	\caption{Linear Landau damping: Snapshots of $p_{2d}(x, v_1, t_{fin})$ produced by HDP and HDP-AS with $\alpha = 0.01$, $N_{\text {eff }} = 5\times 10^{-6}$, $N_{\text {eff }}^F = 1\times 10^{-5}$ and $t_{fin} = 5$.}
	\label{0.01dist}
\end{figure}

\begin{figure}[htbp]
	\centering
	\includegraphics[width=0.45\textwidth]{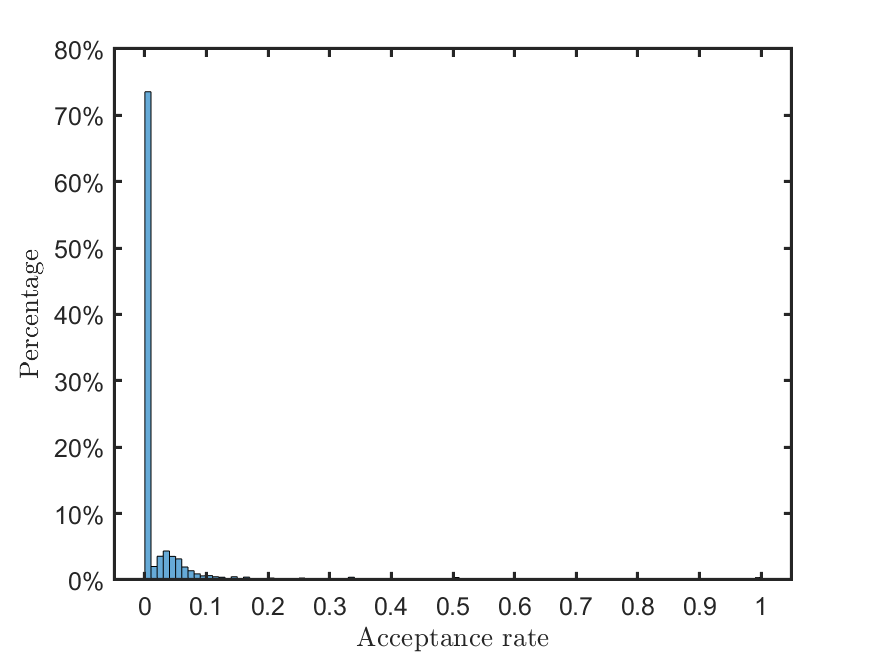}
	\caption{Linear Landau damping with $\alpha = 0.01$, $N_{\text {eff }} = 5\times 10^{-6}$, $N_{\text {eff }}^F = 1\times 10^{-5}$ and $t_{fin} = 5$: Histogram of the acceptance rates in HDP for all grid points $x_k$ and time $t$. Over 73\% of the acceptance rates less than 0.01, and over 95\% of them less than 0.1. Low acceptance rates lead to low efficiency in the acceptance-rejection sampling within the HDP method.}
	\label{linear landau acc rate}
\end{figure}

\begin{figure}[htbp]
	\centering
	\begin{subfigure}[b]{0.45\textwidth}
		\centering
		\includegraphics[width=\linewidth]{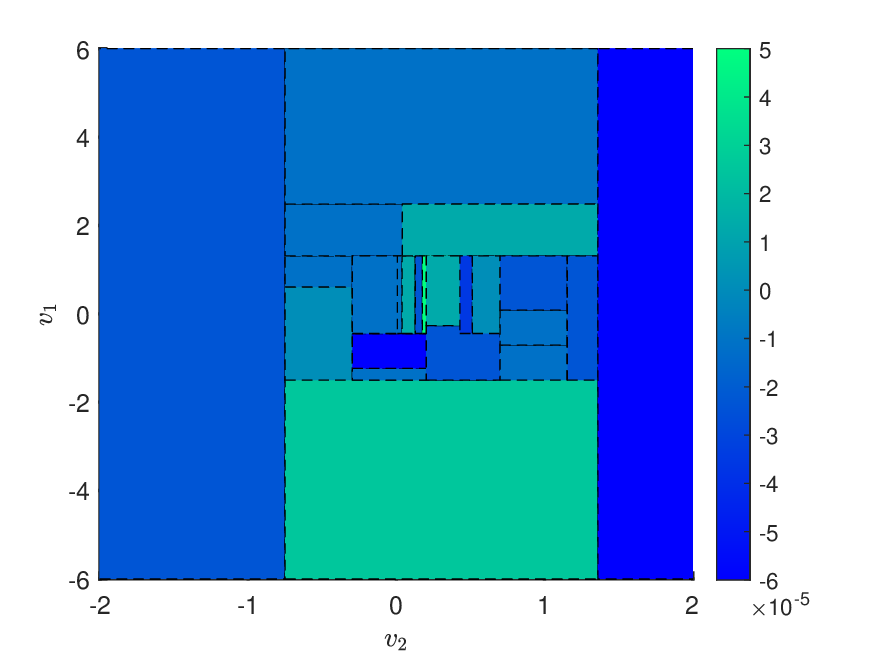}
		\caption{$t = 1$.}
	\end{subfigure}
	\hfill
	\begin{subfigure}[b]{0.45\textwidth}
		\centering
		\includegraphics[width=\linewidth]{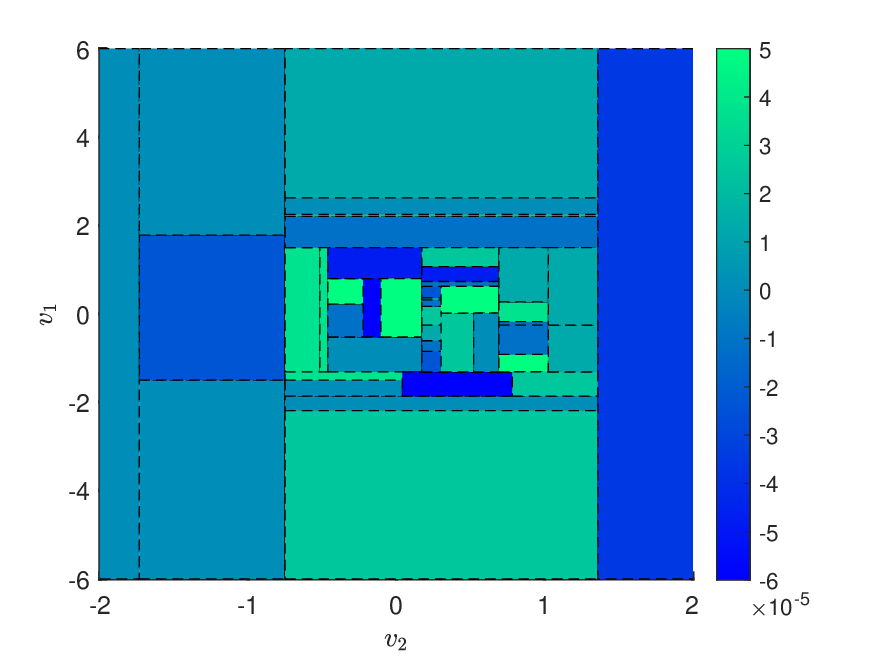}
		\caption{$t = 2$.}
	\end{subfigure}
	\hfill
	\begin{subfigure}[b]{0.45\textwidth}
		\centering
		\includegraphics[width=\linewidth]{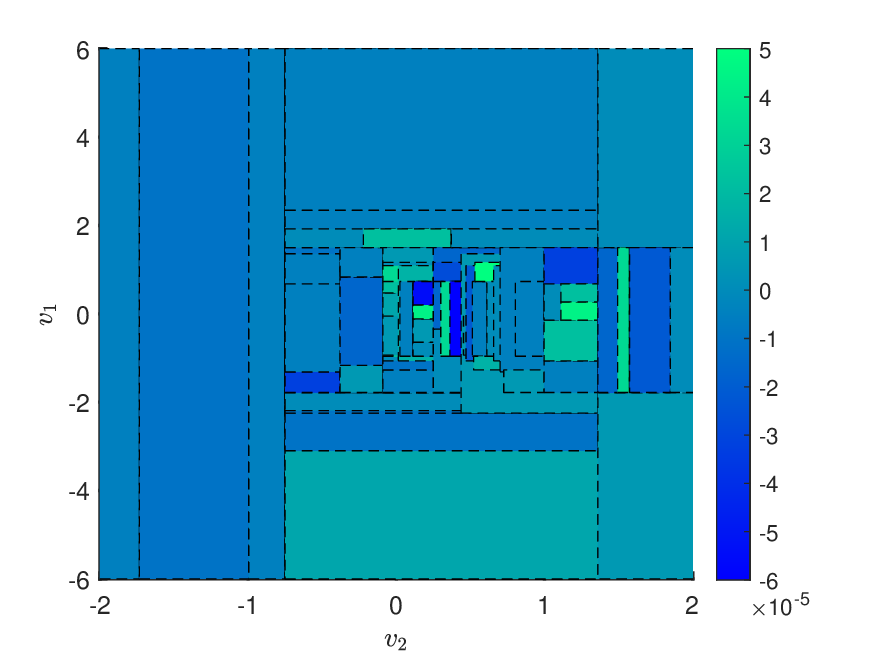}
		\caption{$t = 3$.}
	\end{subfigure}
	\hfill
	\begin{subfigure}[b]{0.45\textwidth}
		\centering
		\includegraphics[width=\linewidth]{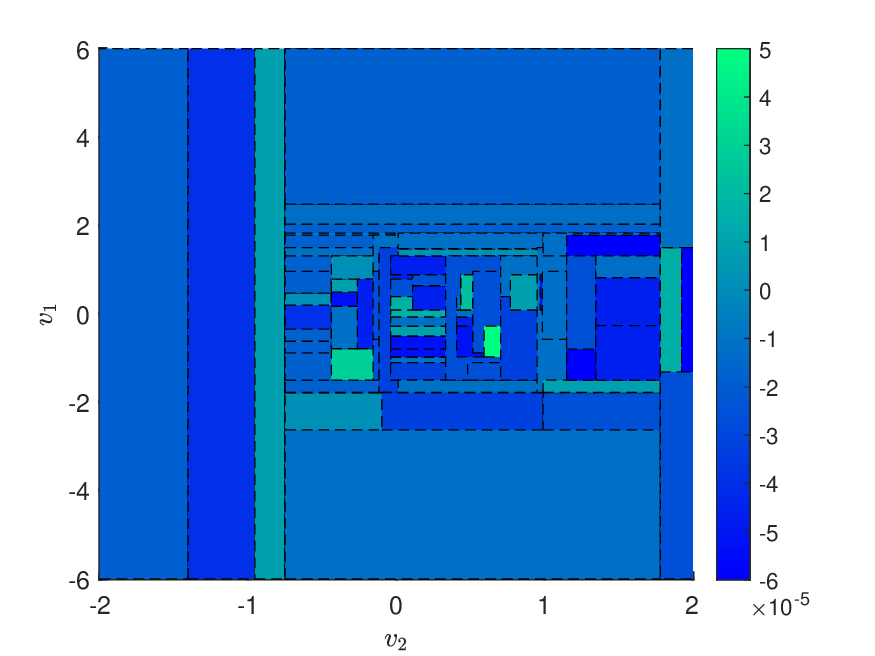}
		\caption{$t = 4$.}
	\end{subfigure}
	\caption{Linear Landau damping with $\alpha = 0.01$, $N_{\text {eff }} = 5\times 10^{-6}$, $N_{\text {eff }}^F = 1\times 10^{-5}$ and $t_{fin} = 5$: The projection of adaptive piecewise constant approximation $\overline{ \Delta m_k}(\bv)$ in $v_1v_2$-plane with $k = 50$ at different instants. In accordance with the changes of the distribution $f(x, \bv, t)$ (see Figure~\ref{0.01dist} for the 2-D projections in $v_1x$-plane), the projection exhibits significant changes in the region $(v_1, v_2)\in \left[-2,2\right]\times \left[-2,2\right]$, and thus we depict the snapshots only for $v_2 \in \left[-2,2\right]$.}
	\label{0.01deltam}
\end{figure}

First we consider the linear Landau damping problem. The initial data are equilibrium distribution $f(x,\bv,0) = m(x,\bv,0)$ with the macroscopic quantities
\begin{equation}\label{landau}
	\left\{\begin{array}{l}
		\rho(x, 0)=1+\alpha \sin (x), \\
		\vec{u}(x, 0)=\vec{0}, \\
		T(x, 0)=1,
	\end{array}\right.
\end{equation}
where the constant $\alpha$ characterizes the amplitude of the perturbation. We conduct the tests until the final time $t_{fin} = 5$. Table \ref{tableaccuracy} presents the efficiency tests for $\alpha  = 0.001, 0.01, 0.1$ and different sample sizes. The HDP-AS method runs roughly ten times faster than the HDP method, along with smaller errors. For example, when $\alpha = 0.01$, $N_{\text {eff }} = 5\times 10^{-6}$ and $N_{\text {eff }}^F = 1\times 10^{-5}$, HDP-AS takes 403.26 seconds while HDP takes 4626.91 seconds, achieving an acceleration of 11.47 times. In terms of the accuracy, the relative $L^2$ error $\mathcal{E}_2[p_{2d}](t_{fin})$ of HDP-AS (0.0264) is also preferable to that of HDP (0.0451). Figure~\ref{0.01dist} displays the snapshots of $p_{2d}(x, v_1, t_{fin})$ produced by HDP and HDP-AS corresponding to this group of parameters and as excepted, a good agreement can be readily observed. Figure~\ref{linear landau acc rate} displays the histogram of the acceptance rates in HDP. More than 73\% of the acceptance rates are less than 0.01. The low acceptance rate leads to a high sampling cost. On the contrary, HDP-AS achieves high-dimensional sampling without rejection by approximating the source term $\Delta m_k(\bv)$ with an adaptive piecewise constant function $\overline{ \Delta m_k}(\bv)$ defined in Eq.~\eqref{ada pie con}. Figure \ref{0.01deltam} plots $\int_{\mathbb{R}^3}\overline{ \Delta m_k}(\bv) \D v_3$, the projection of adaptive piecewise constant approximation $\overline{ \Delta m_k}(\bv)$ in $v_1v_2$-plane with $k = 50$. In accordance with the changes of the distribution $f(x, \bv, t)$ (see Figure~\ref{0.01dist} for the 2-D projections in $v_1x$-plane), this projection exhibits significant changes in the region $(v_1, v_2)\in \left[-2,2\right]\times \left[-2,2\right]$, which reflects the intrinsic adaptivity of HDP-AS. 
Just because of the adaptivity of $\overline{ \Delta m_k}(\bv)$,  
new deviational particles sampled by HDP-AS are of higher quality than those by HDP and thus able to capture the characteristics of $\Delta m_k(\bv)$ accurately with a smaller size. 
For instance, 
HDP-AS samples $177$ new P-particles and $166$ new N-particles, whereas HDP samples $412$ new P-particles and $370$ new N-particles
for $\alpha = 0.01$, $N_{\text {eff }} = 5\times 10^{-6}$, $N_{\text {eff }}^F = 1\times 10^{-5}$ and $t = 1$ in this example. 
Much less new deviational particles being added may enable other remaining steps such as simulating collisions via TA and advection to take less time (see Figure~\ref{ratio}).





\begin{table}[htbp]
	\centering
	\begin{tabular}{ccccccc}
		\hline
		\multirow{2}{*}{$\alpha$} & \multirow{2}{*}{$N_{\text {eff }}$} & \multirow{2}{*}{$N_{\text {eff }}^F$} &\multicolumn{2}{c}{HDP} &\multicolumn{2}{c}{HDP-AS}\\ \cline{4-7} 
		&    &   &    $\mathcal{E}_2[p_{2d}](t_{fin})$ & Time/s &$\mathcal{E}_2[p_{2d}](t_{fin})$ & Time/s \\ 
		\midrule
		&$2\times10^{-5}$  &  $4\times10^{-5}$      & 0.1379 & 1951.41   & 0.0523 & 97.79  \\ 
		0.001 & $1\times10^{-5}$  &$2\times10^{-5}$    & 0.0829& 2194.63 & 0.0306 & 183.39  \\ 
		&$5\times10^{-6}$  &  $1\times10^{-5}$    & 0.0324 & 2692.26 & 0.0228 & 359.10   \\ \midrule
		&$2\times10^{-5}$  & $4\times10^{-5}$  & 0.1679 & 3049.58 & 0.0577 & 111.58\\
		0.01 & $1\times10^{-5}$  & $2\times10^{-5}$  &  0.0974 & 3833.22 &0.0381 & 216.10  \\ 
		& $5\times10^{-6}$  & $1\times10^{-5}$    & 0.0451 & 4626.91  & 0.0264 & 403.26   \\ \midrule
		& $2\times10^{-5}$  &    $2\times10^{-5}$    & 0.2020 & 7281.12  & 0.0999 & 345.94  \\ 
		
		0.1& $1\times10^{-5}$  & $1\times10^{-5}$      & 0.1275 & 7736.82  & 0.0688 & 626.06     \\ 
		& $5\times10^{-6}$  &   $5\times10^{-6}$      & 0.0681 & 12435.64  & 0.0584 & 1380.85   \\
		\hline 
	\end{tabular}
	\caption{Linear Landau damping: Total wall time and the relative $L^2$ error $\mathcal{E}_2[p_{2d}](t_{fin})$ (see Eq.~\eqref{err}) under different $\alpha$ and sample sizes. HDP-AS attains an acceleration of approximately ten times against HDP and maintains smaller errors.}
	\label{tableaccuracy}
\end{table}


\subsection{Nonlinear Landau damping}
\par
\begin{figure}[htbp]
	\centering
	\begin{subfigure}[b]{0.45\textwidth}
		\centering
		\includegraphics[width=\linewidth]{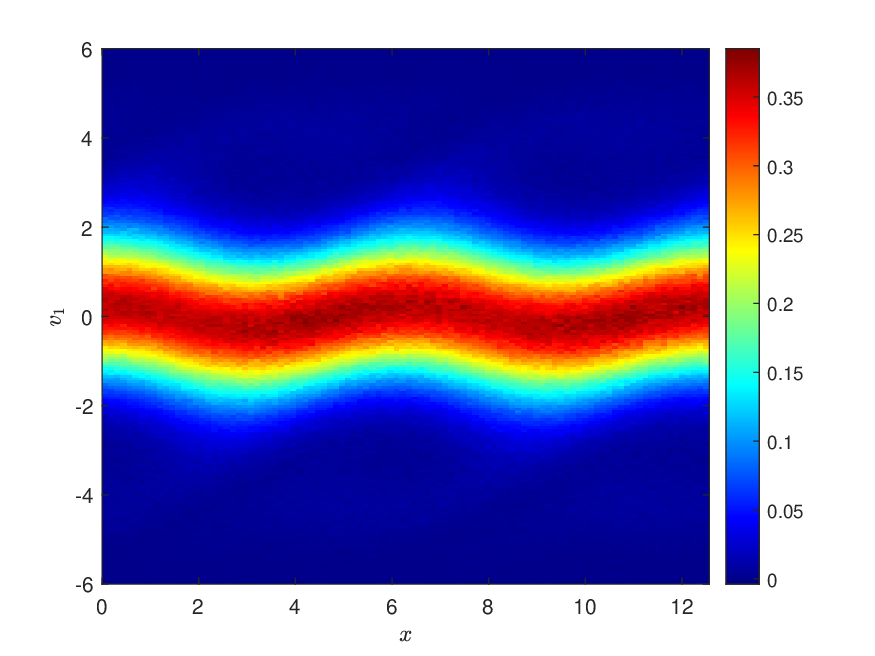}
		\caption{HDP.}
	\end{subfigure}
	\hfill
	\begin{subfigure}[b]{0.45\textwidth}
		\centering
		\includegraphics[width=\linewidth]{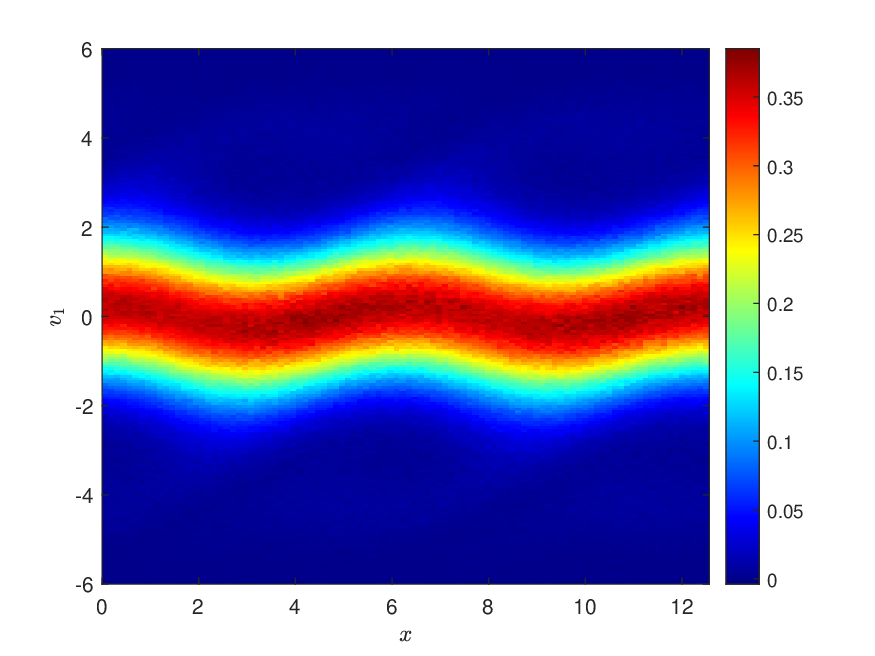}
		\caption{HDP-AS.}
	\end{subfigure}
	\caption{Nonlinear Landau damping: Snapshots of $p_{2d}(x, v_1, t_{fin})$ produced by HDP and HDP-AS with $\alpha = 0.4$, $N_{\text {eff }} = 1\times 10^{-6}$, $N_{\text {eff }}^F = 1\times 10^{-6}$ and $t_{fin} = 5$.}
	\label{nonlinear landau dist}
\end{figure}

\begin{figure}[htbp]
	\centering
	\includegraphics[width=0.45\linewidth]{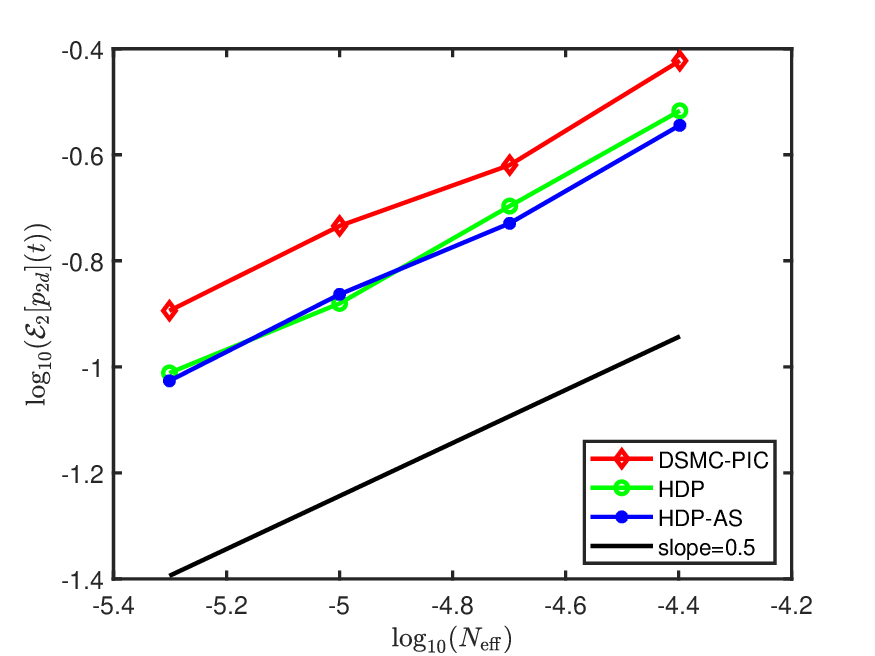}
	\caption{Nonlinear Landau damping with $\alpha = 0.4$ and $t_{fin} = 0.5$: The HDP, HDP-AS and DSMC-PIC methods all exhibit a half order convergence. When the sample sizes are identical, the errors of both HDP and HDP-AS are smaller than that of DSMC-PIC.}
	\label{convergence}
\end{figure}

\begin{figure}[htbp]
	\centering
	\includegraphics[width=0.45\textwidth]{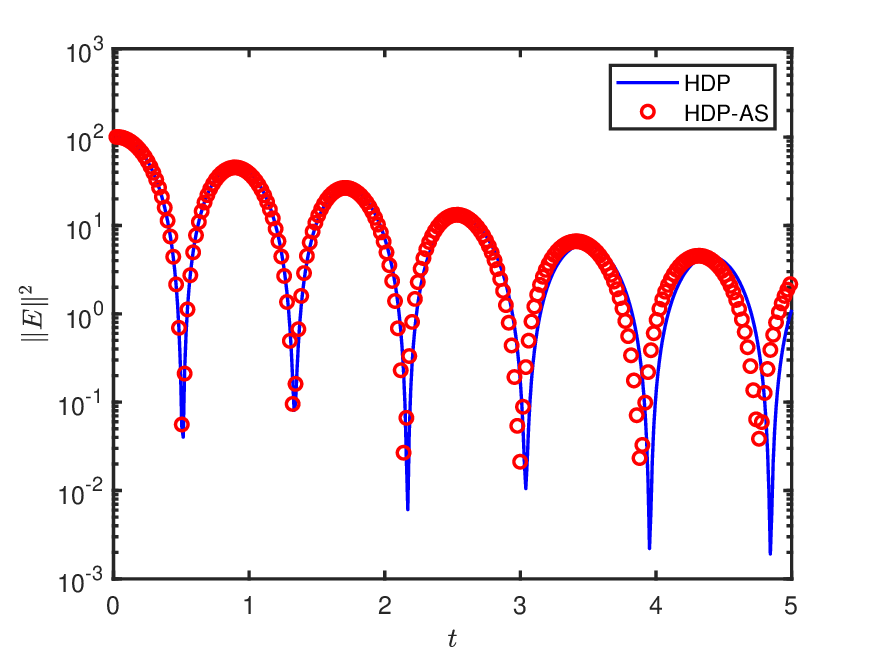}
	\caption{Nonlinear Landau damping with $\alpha = 0.4$, $N_{\text {eff }} = 1\times 10^{-6}$, $N_{\text {eff }}^F = 1\times 10^{-6}$ and $t_{fin} = 5$: Comparison of the decay of energy $\Vert E\Vert^2$ between HDP and HDP-AS.}
	\label{nonlinear landau phenomenon}
\end{figure}

 We consider the nonlinear Landau damping problem with $\alpha = 0.4$ in Eq.~\eqref{landau}. Figure \ref{nonlinear landau dist} illustrates the good agreement of solutions $p_{2d}(x, v_1, 5)$ between the HDP method and the HDP-AS method. Figure \ref{convergence} presents the convergence tests and all methods show a half order convergence. Furthermore, both HDP and HDP-AS exhibit smaller errors than DSMC-PIC after taking the same sample size. 
For the Landau damping problem, the damping of electric field energy
\begin{equation}\label{e}
	\Vert E\Vert^2 = \Delta x \sum_{k=1}^{N_x} \left(E_k\right)^2
\end{equation}
is also be studied, where $E_k$ is the value of the electric field $E$ at the position $x_k$. Figure \ref{nonlinear landau phenomenon} shows clearly the Landau damping phenomenon, and the curve produced by HDP-AS agrees well with that by HDP.


\subsection{Two stream instability}
\par
\begin{figure}[htbp]
	\centering
	\begin{subfigure}[b]{0.45\textwidth}
		\centering
		\includegraphics[width=\linewidth]{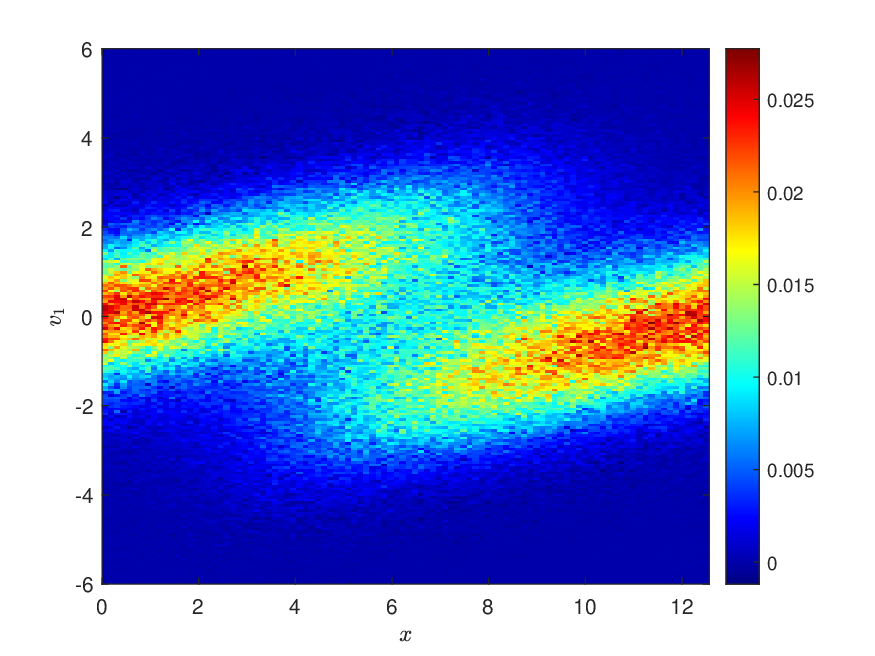}
		\caption{HDP, $t = 2$.}
	\end{subfigure}
	\hfill
	\begin{subfigure}[b]{0.45\textwidth}
		\centering
		\includegraphics[width=\linewidth]{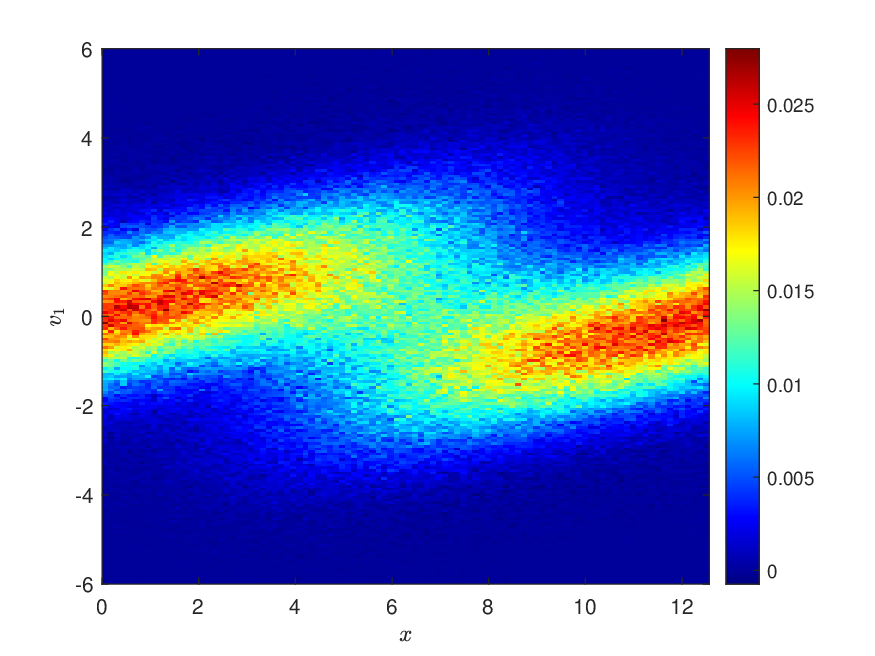}
		\caption{HDP-AS, $t = 2$.}
	\end{subfigure}
	\hfill
	\begin{subfigure}[b]{0.45\textwidth}
		\centering
		\includegraphics[width=\linewidth]{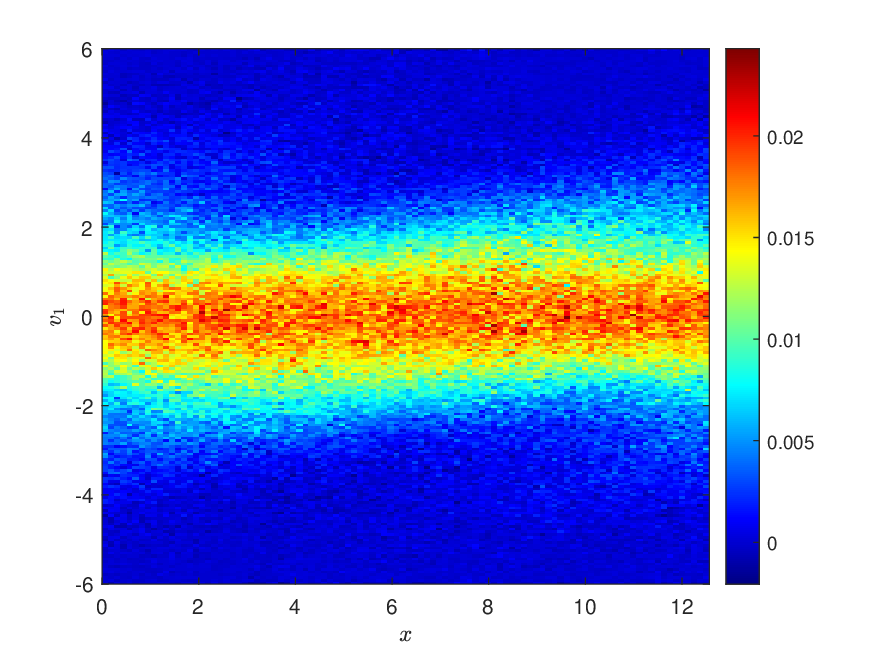}
		\caption{HDP, $t = 4$.}
	\end{subfigure}
	\hfill
	\begin{subfigure}[b]{0.45\textwidth}
		\centering
		\includegraphics[width=\linewidth]{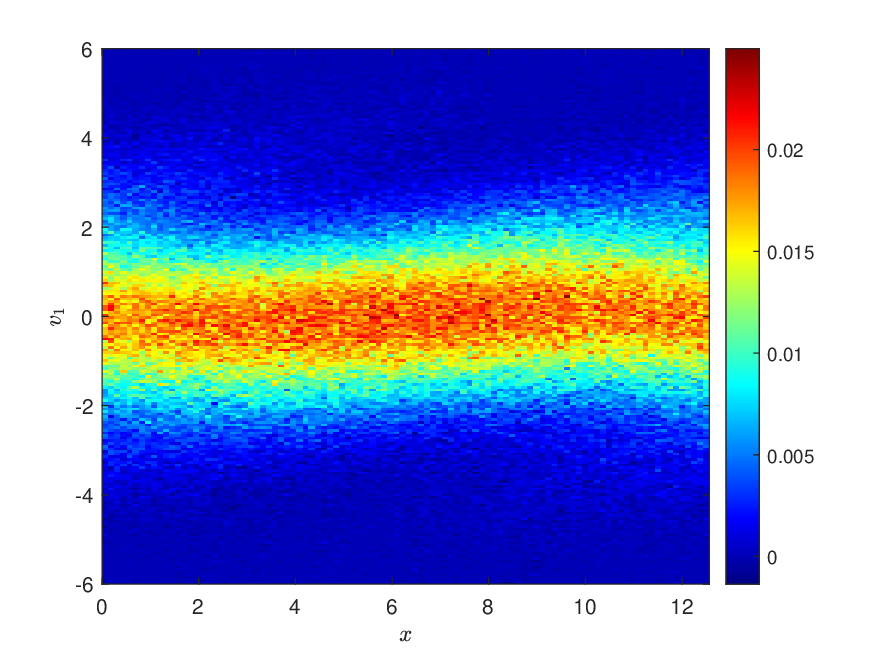}
		\caption{HDP-AS, $t = 4$.}
	\end{subfigure}
	\hfill
	\begin{subfigure}[b]{0.45\textwidth}
		\centering
		\includegraphics[width=\linewidth]{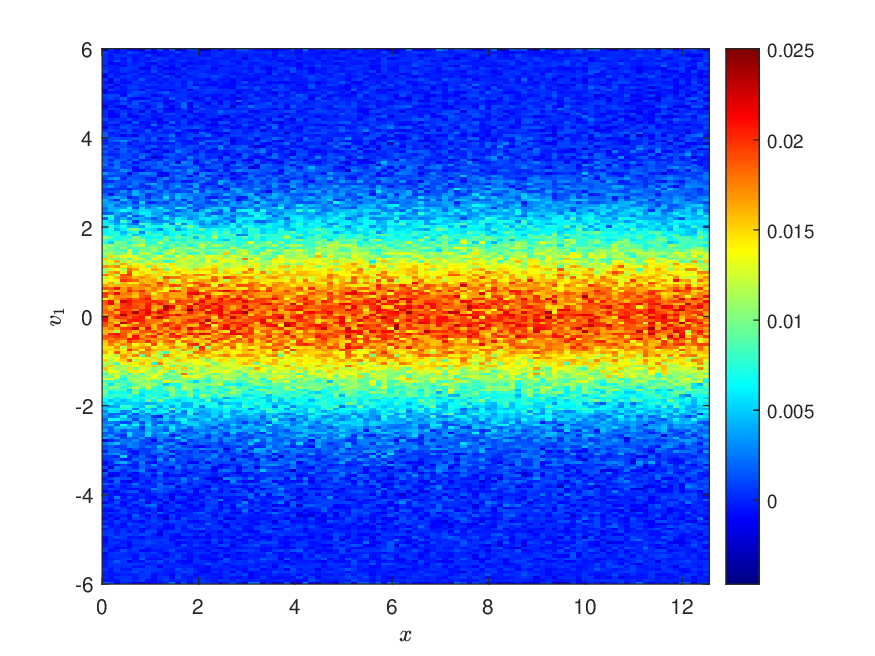}
		\caption{HDP, $t = 8$.}
	\end{subfigure}
	\hfill
	\begin{subfigure}[b]{0.45\textwidth}
		\centering
		\includegraphics[width=\linewidth]{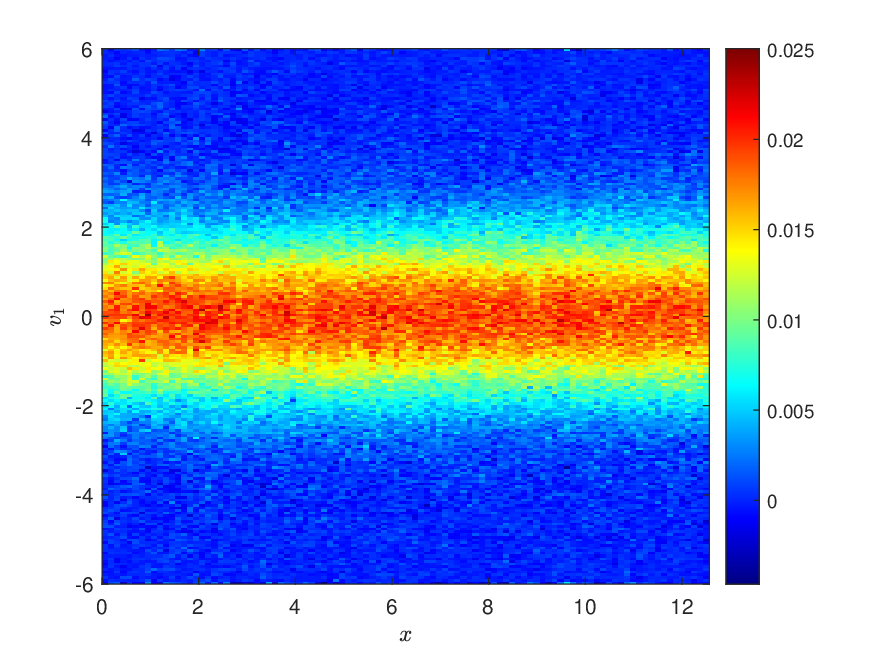}
		\caption{HDP-AS, $t = 8$.}
	\end{subfigure}
	\caption{Two stream instability: Plots of $p_{2d}(x, v_1, t)$ at $t = 2,4,8$ for the HDP and HDP-AS solutions with $N_{\text {eff }} = 1\times 10^{-6}$ and $N_{\text {eff }}^F = 1\times 10^{-6}$.}
	\label{TS figure}
\end{figure}

\begin{figure}[htbp]
	\centering
	\begin{subfigure}[b]{0.45\textwidth}
		\centering
		\includegraphics[width=\linewidth]{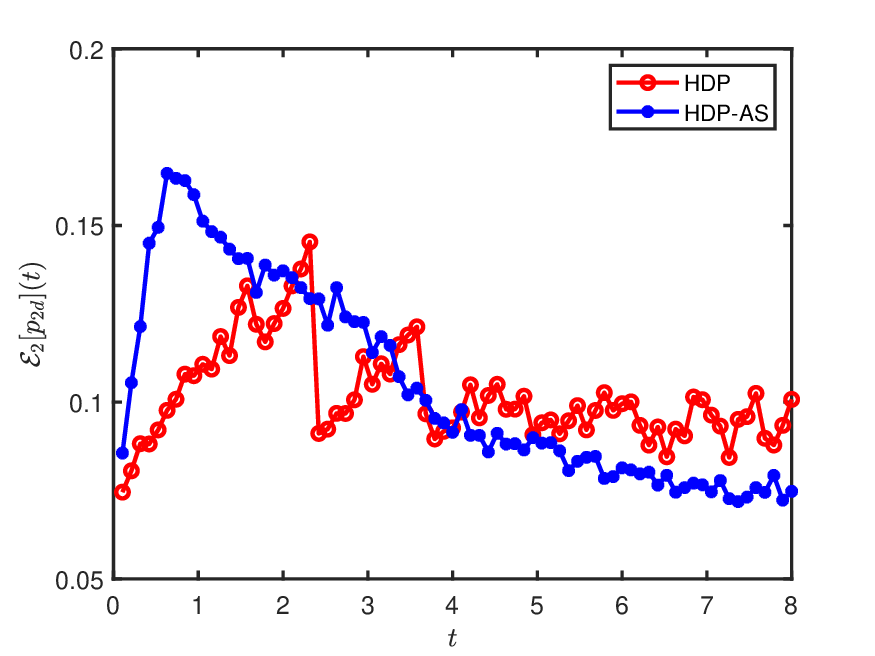}
		\caption{$\mathcal{E}_2[p_{2d}](t)$.}
	\end{subfigure}
	\hfill
	\begin{subfigure}[b]{0.45\textwidth}
		\centering
		\includegraphics[width=\linewidth]{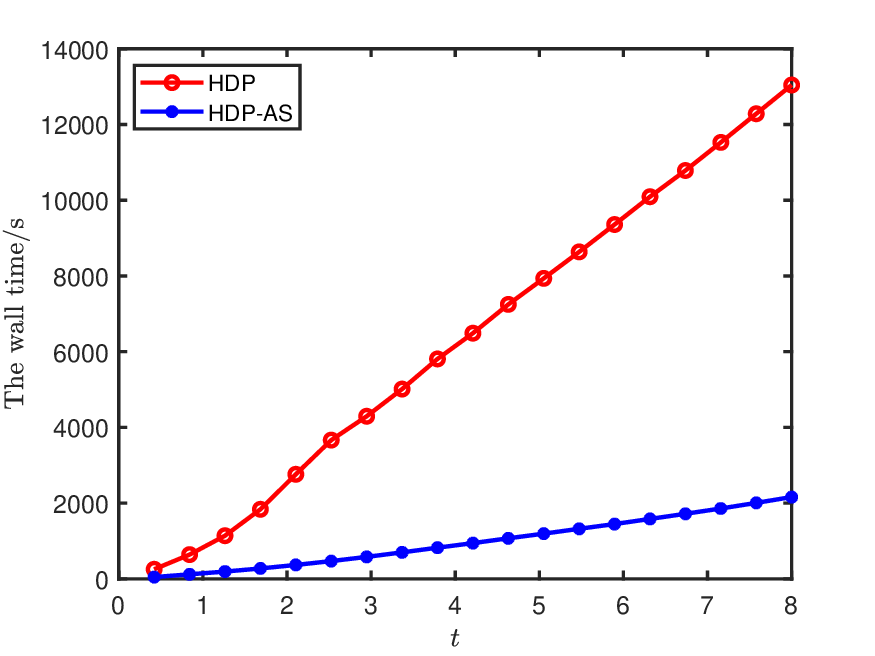}
		\caption{The wall time.}
	\end{subfigure}
	\caption{Two stream instability: History of the relative $L^2$ error $\mathcal{E}_2[p_{2d}](t)$ and the wall time for HDP and HDP-AS with $N_{\text {eff }} = 1\times 10^{-6}, N_{\text {eff }}^F = 1\times 10^{-6}$. The error evolutions of the two methods are comparable, yet the wall time of HDP-AS is significantly less than that of HDP.}
	\label{TS error cost}
\end{figure}

 In this numerical experiment, the initial data are
\begin{equation}
	f\left(x, \bv,0\right)=\frac{1}{12 \pi}(1+0.5 \cos (0.5 x)) \exp\left(-\frac{v_1^2+v_2^2+v_3^2}{2}\right)\left(1+5 v_1^2\right),
\end{equation}
and we run the tests up to the final time $t_{fin} = 8$. Figure \ref{TS figure} presents the snapshots of $p_{2d}(x, v_1, t)$ at different instants. The two stream is observed when $t = 2$, and the distribution $p_{2d}(x, v_1, t)$ approaches the equilibrium state for $t\geq 4$ due to the effect of collisions. Figure \ref{TS error cost} exhibits the relative $L^2$ error $\mathcal{E}_2[p_{2d}](t)$ and the wall time throughout the simulation process. Up to $t_{fin} = 8$, HDP-AS  takes 2190 seconds whereas HDP needs 13866 seconds. In a word, HDP-AS significantly reduces the wall time compared to HDP while maintaining the same accuracy.

\subsection{Bump on tail}
\par
\begin{figure}[htbp]
	\centering
	\begin{subfigure}[b]{0.49\textwidth}
		\centering
		\includegraphics[width=\linewidth]{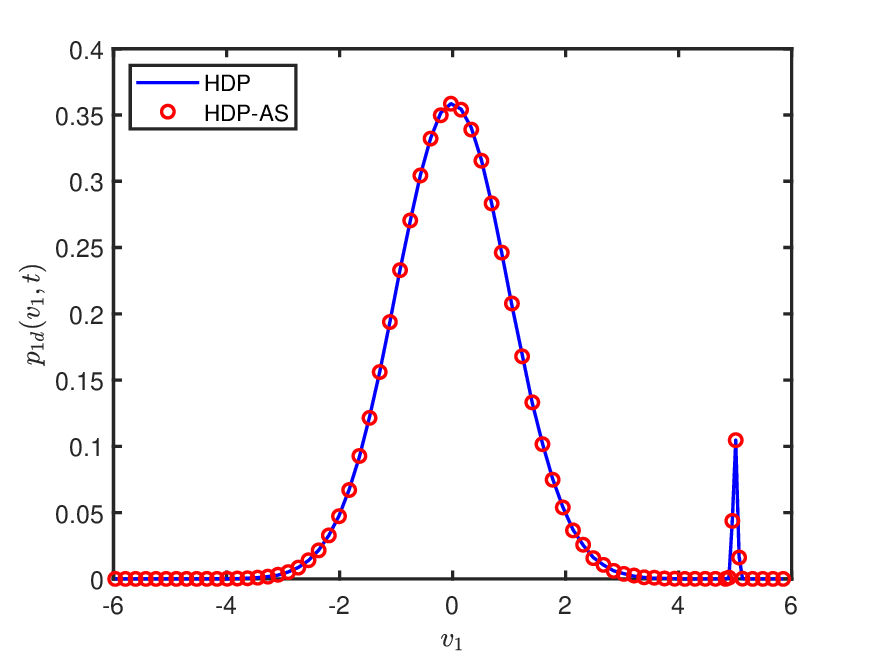}
		\caption{$t = 0.0$.}
	\end{subfigure}
	\hfill
	\begin{subfigure}[b]{0.49\textwidth}
		\centering
		\includegraphics[width=\linewidth]{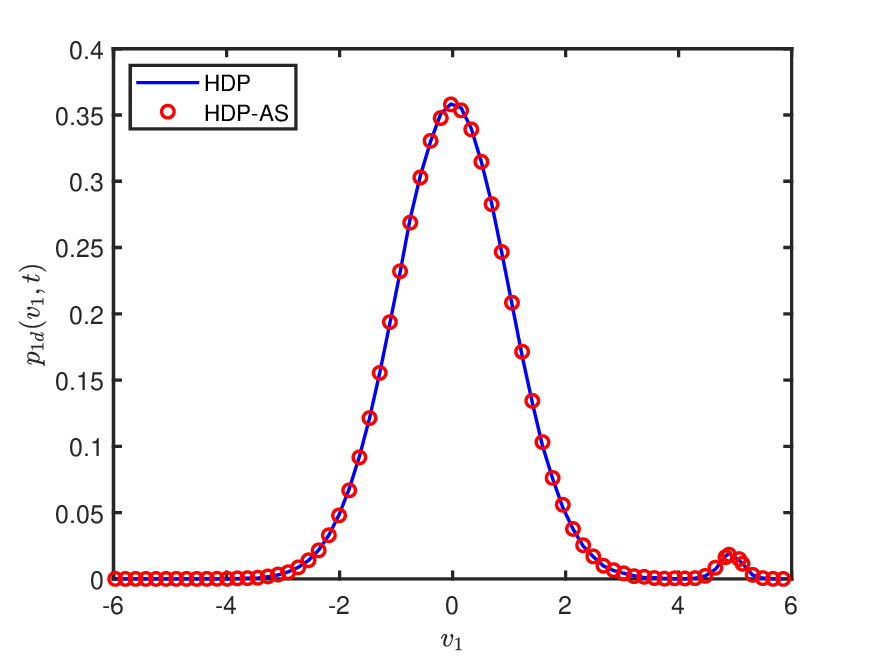}
		\caption{$t = 0.5$.}
	\end{subfigure}
	\hfill
	\begin{subfigure}[b]{0.49\textwidth}
		\centering
		\includegraphics[width=\linewidth]{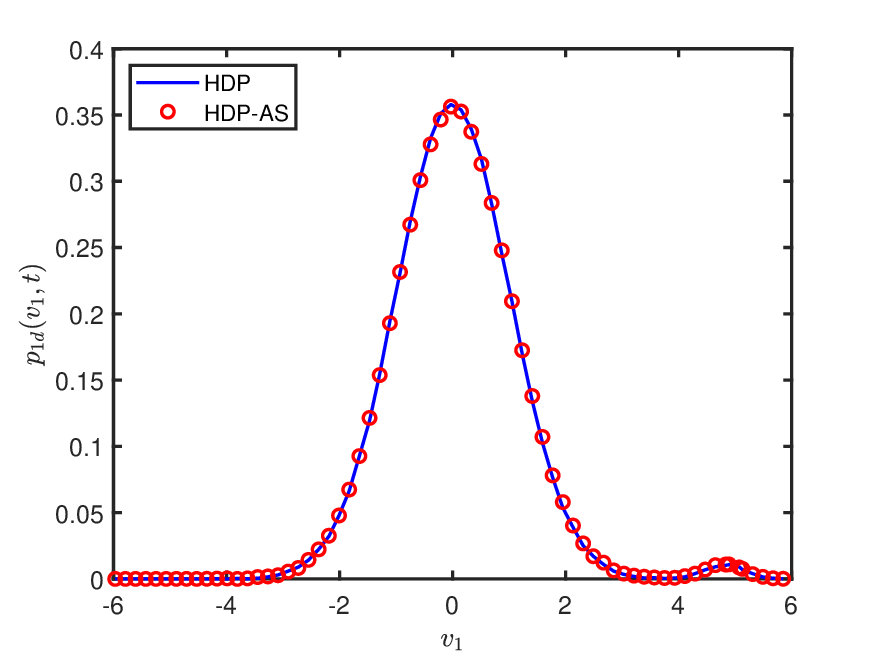}
		\caption{$t = 1.0$.}
	\end{subfigure}
	\hfill
	\begin{subfigure}[b]{0.49\textwidth}
		\centering
		\includegraphics[width=\linewidth]{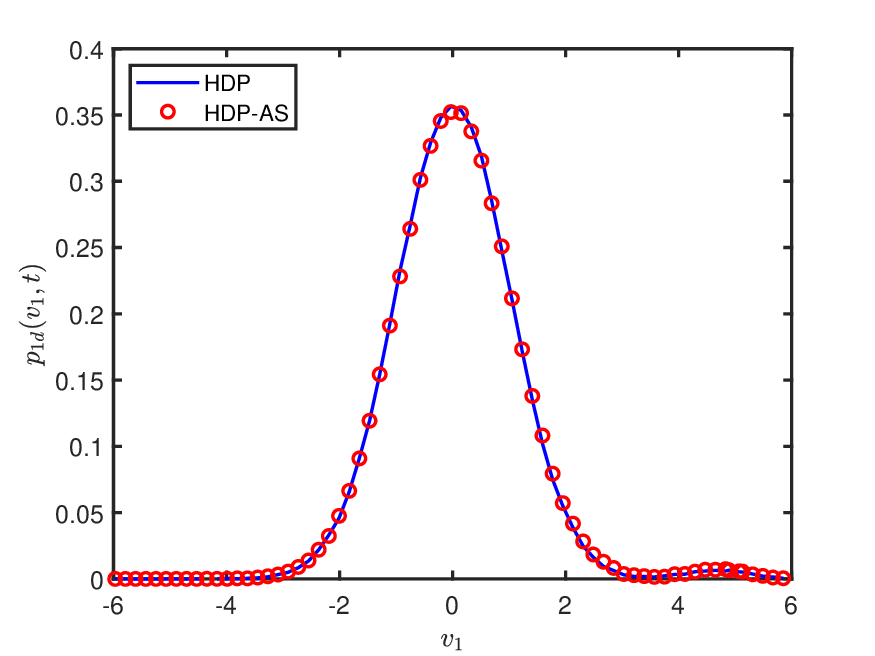}
		\caption{$t = 2.0$.}
	\end{subfigure}
	\caption{Bump on tail: Time evolutions of $p_{1d}(v_1, t)$ for the HDP and HDP-AS methods with $N_{\text {eff }} = 1\times 10^{-6}, N_{\text {eff }}^F = 1\times 10^{-6}, \nu = 0.1$ and $q = 16$.}
	\label{bot figure}
\end{figure}

 In this case, we test the bump on tail problem, whose initial data contain a central Maxwellian and a small bump at high energy,
\begin{equation}
	\begin{aligned}
		f(x,\bv,0) =& \left[\frac{\beta}{(2\pi T_a)^{3/2}}\exp\left(-\frac{|\bv|^2}{2 T_a}\right) + \frac{(1-\beta)}{(2\pi T_b)^{3/2}}\exp\left(-\frac{|\bv-\vec{u}_b|^2}{2 T_b}\right)\right] \\
		\times&\frac{1}{(2\pi T_x)^{1/2}} \exp\left(-\frac{(x-x_c)^2}{2T_x}\right),
	\end{aligned}
\end{equation}
where $\beta = 0.9, T_a = 1, T_b = 0.001, \vec{u}_b = \left[5,0,0\right], x_c = 2\pi, T_x = 0.25$. Here we set the final time $t_{fin} = 2$. Figure \ref{bot figure} plots the HDP-AS solutions against the HDP ones. We are able to observe there that the high energy part decays and shifts towards the Maxwellian state for both solutions. Table \ref{BOT nu q} shows the sensitivity tests about parameters $\nu$ and $q$ needed in Algorithm \ref{adaptive alg}. The parameter $\nu$ defined in Eq.~\eqref{stop criter} adjusts the depth of partition and the parameter $q$ defined in Eq.~\eqref{split c} is related to the number of selectable partition points. Numerical results in Table \ref{BOT nu q} demonstrates that HDP-AS is insensitive to these two parameters. The errors $\mathcal{E}_2[p_{2d}](t_{fin})$ of the nine groups of data are all around 0.12, which is close to the error 0.1378 of HDP. The wall time of HDP-AS slightly increases with the decrease of $\nu$ and the increase of $q$, because a smaller $\nu$ means a finer partition and a larger $q$ means more selectable partition points to be examined. On the whole, the wall time of HDP-AS is around 300 seconds, significantly lower than 2414.43 seconds of HDP.
\begin{table}[htbp]
	\centering
	\begin{tabular}{ccccc}
		\hline
		Method                & $\nu$                    & $q$  & $\mathcal{E}_2[p_{2d}](t_{fin})$ & Time/s   \\ \hline
		HDP                    & -                     & -  & 0.1378 & 2414.43 \\ \hline
		\multirow{9}{*}{HDP-AS} & \multirow{3}{*}{0.1}  & 4  & 0.1117 & 243.57  \\ \cline{3-5} 
		&                       & 16 & 0.1150  & 247.36  \\ \cline{3-5} 
		&                       & 64 & 0.1160  & 258.94  \\ \cline{2-5} 
		& \multirow{3}{*}{0.05} & 4  & 0.1225 & 244.41  \\ \cline{3-5} 
		&                       & 16 & 0.1073 & 248.30  \\ \cline{3-5} 
		&                       & 64 & 0.1161 & 261.08  \\ \cline{2-5} 
		& \multirow{3}{*}{0.01} & 4  & 0.1250  & 359.29  \\ \cline{3-5} 
		&                       & 16 & 0.1209 & 364.07  \\ \cline{3-5} 
		&                       & 64 & 0.1087 & 385.95  \\ \hline
	\end{tabular}
	\caption{Bump on tail: Sensitivity tests about the parameters $\nu$ and $q$ in Algorithm \ref{adaptive alg} with $N_{\text {eff }} = 1\times 10^{-6}$ and $N_{\text {eff }}^F = 1\times 10^{-6}$. It can be readily seen that HDP-AS is not sensitive to $\nu$ and $q$, and runs approximately ten times faster than HDP while keeping the same accuracy. }
	\label{BOT nu q}
\end{table}


\begin{figure}[htbp]
	\centering
	\begin{subfigure}[b]{0.49\textwidth}
		\centering
		\includegraphics[width=\linewidth]{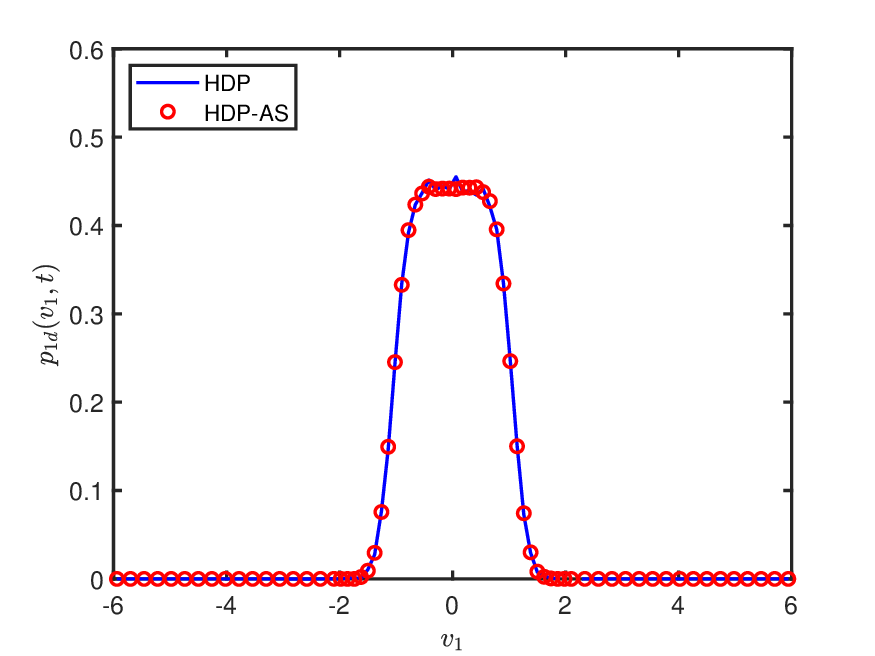}
		\caption{$t = 0.0$.}
	\end{subfigure}
	\hfill
	\begin{subfigure}[b]{0.49\textwidth}
		\centering
		\includegraphics[width=\linewidth]{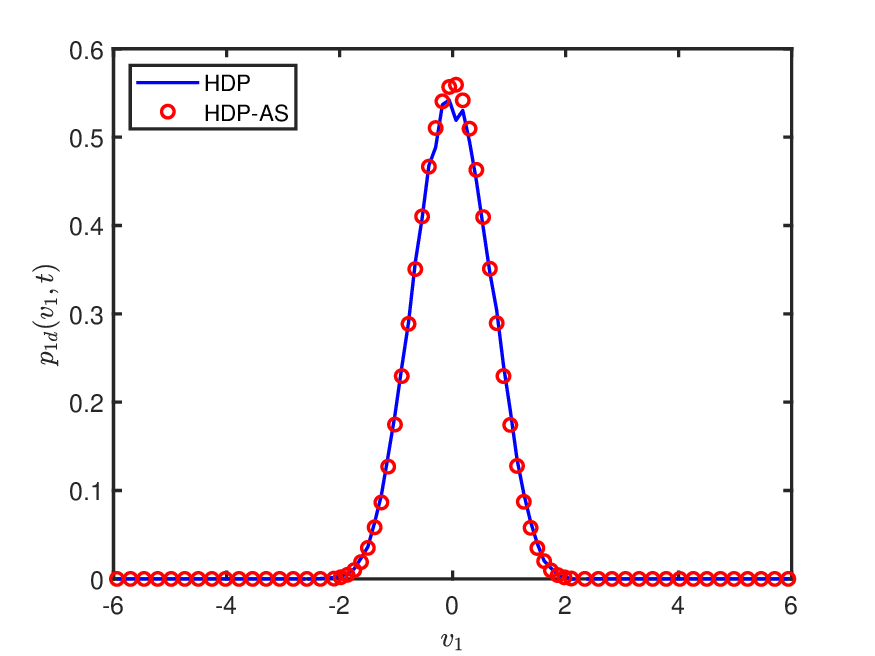}
		\caption{$t = 0.25$.}
	\end{subfigure}
	\hfill
	\begin{subfigure}[b]{0.49\textwidth}
		\centering
		\includegraphics[width=\linewidth]{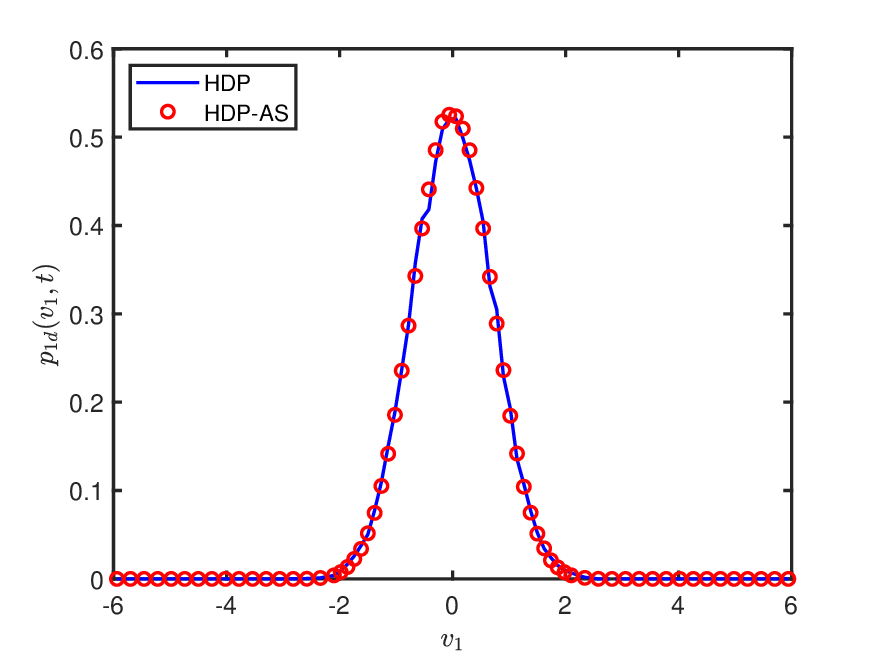}
		\caption{$t = 0.5$.}
	\end{subfigure}
	\hfill
	\begin{subfigure}[b]{0.49\textwidth}
		\centering
		\includegraphics[width=\linewidth]{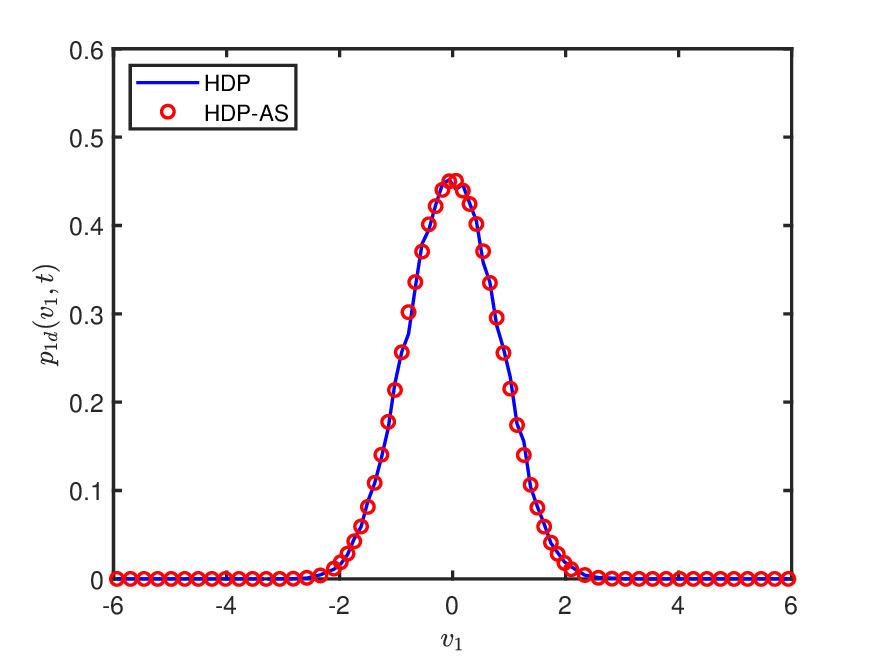}
		\caption{$t = 1.0$.}
	\end{subfigure}
	\caption{Rosenbluth's test problem: Time evolutions of $p_{1d}(v_1, t)$ for HDP  and HDP-AS  with $N_{\text {eff }} = 1\times 10^{-6}, N_{\text {eff }}^F = 5\times 10^{-7}, \nu = 0.1$ and $q = 16$.}
	\label{Ros figure}
\end{figure}

\begin{table}[htbp]
	\centering
	\begin{tabular}{ccccc}
		\hline
		Method                 & $\nu$                    & $q$  & $\mathcal{E}_2[p_{2d}](t_{fin})$ & Time/s   \\ \hline
		HDP                    & -                     & -  & 0.0965 & 3946.71 \\ \hline
		\multirow{9}{*}{HDP-AS} & \multirow{3}{*}{0.1}  & 4  & 0.0225 & 191.91  \\ \cline{3-5} 
		&                       & 16 & 0.0225 & 193.08  \\ \cline{3-5} 
		&                       & 64 & 0.0220  & 195.86  \\ \cline{2-5} 
		& \multirow{3}{*}{0.05} & 4  & 0.0217 & 193.52  \\ \cline{3-5} 
		&                       & 16 & 0.0217 & 194.13  \\ \cline{3-5} 
		&                       & 64 & 0.020 & 197.58  \\ \cline{2-5} 
		& \multirow{3}{*}{0.01} & 4  & 0.0230  & 285.12  \\ \cline{3-5} 
		&                       & 16 & 0.0223 & 290.17  \\ \cline{3-5} 
		&                       & 64 & 0.0233 & 385.95  \\ \hline
	\end{tabular}
	\caption{Rosenbluth's test problem: Sensitivity tests about the parameters $\nu$ and $q$ in Algorithm \ref{adaptive alg} on the HDP-AS method with $N_{\text {eff }} = 1\times 10^{-6}$ and $N_{\text {eff }}^F = 5\times 10^{-7}$.}
	\label{Ros nu q}
\end{table}

\subsection{Rosenbluth's test problem}
\par Finally we examine the Rosenbluth's test problem and the initial data are given by 
\begin{equation}
	f(x,\bv,0) = 0.01 \exp(-10(|\bv|-1)^2)(1+0.5\sin(0.5x)).
\end{equation}
Here we adopt the final time $t_{fin}$ as 1. Figure \ref{Ros figure} illustrates the good agreement between the solutions obtained from HDP-AS and HDP at different instants. Table \ref{Ros nu q} presents the sensitivity tests for Rosenbluth's test problem. HDP-AS method is insensitive to the parameters and it runs approximately ten times (191.91-385.95 seconds) faster than the HDP method (3946.71 seconds) with smaller errors.

\section{Conclusion and discussion}
\label{sec:con}

The main contributions of this work include two parts:

\begin{itemize}
	
	\item Proposing an alternative adaptive strategy for sampling from $\Delta m_k(\bv)_{\pm}$ in the HDP method. It constructs an adaptive piecewise constant approximation of $\Delta m_k(\bv)_{\pm}$ at first, and then the sampling is conducted directly from the piecewise constant function without rejection.
	
	\item Employing the mixture discrepancy in place of the star discrepancy as the uniformity measure accelerates the 
	reconstruction of the adaptive piecewise constant approximation in \cite{Li2016}. The mixture discrepancy has an explicit expression and no external solver is required.
	
\end{itemize}
\par Owing to the above two improvements, the proposed HDP-AS method successfully reduces the computational cost of the HDP method. The same idea could be applied to other sampling tasks, provided that each part of the function to be sampled has already been represented by multiple particle sets (e.g., in this work, the $Q(f_p,m)$ and $Q(f_n,m)$ in $\Delta m_k(\bv)_{\pm}$ are represented by particle sets $\mathbb{M}^{+}_k$ and $\mathbb{M}^{-}_k$ in Eq.~\eqref{Wk}).
There remain some aspects to be explored, such as  
the error analysis of HDP-AS, the application to practical problems in higher dimensions, 
and whether more efficient uniformity measure manner can further accelerate the reconstruction of the adaptive piecewise constant approximation.


\section*{Acknowledgement}
This research was supported by the National Natural Science Foundation of China (Nos.~12325112, 12288101) and the High-performance Computing Platform of Peking University. The authors are sincerely grateful to Dr.~Bokai Yan, the author of \cite{Yan2016}, for providing us the source code of HDP, which greatly facilitates our implementation of HDP-AS as well as  the comparison between HDP and HDP-AS.



\end{document}